\documentclass[twocolumn,aps,pre,floatfix,superscriptaddress]{revtex4-1}
\usepackage{graphicx}
\usepackage{epic,eepic}
\usepackage{graphics}
\usepackage{epsfig}
\usepackage{subfigure}

\newcommand{\be}{\begin{equation}}
\newcommand{\ee}{\end{equation}}
\newcommand{\bea}{\begin{eqnarray}}
\newcommand{\eea}{\end{eqnarray}}

\usepackage{amsmath}
\usepackage{amsfonts}
\usepackage{amssymb}
\usepackage{bm}
\usepackage{color}
\usepackage{graphicx}

\begin{document}
\title{Unravelling the role of phoretic and hydrodynamic interactions in active colloidal suspensions}
\author{A. Scagliarini} \affiliation{IAC-CNR, Isituto per le Applicazioni del Calcolo ``Mauro Picone'', Via dei Taurini 19, 00185 Rome, Italy. E-mail: andrea.scagliarini@cnr.it}
\author{I. Pagonabarraga}\affiliation{CECAM, Centre Europ\'een de Calcul Atomique et Mol\'eculaire, Ecole Polytechnique  F\'ed\'erale  de  Lausanne,  Batochimie, Avenue Forel 2, 1015 Lausanne, Switzerland} 

\begin{abstract}
Active fluids comprise a variety of systems composed of elements immersed in a
fluid environment which can convert some form of energy into
directed motion; as such
they are intrinsically out-of-equilibrium in the absence of any external forcing.
A fundamental problem in the physics of active matter concerns the understanding of 
how the characteristics of the autonomous propulsion and agent-agent interactions determine 
the collective dynamics of the system.
We study numerically suspensions of self-propelled diffusiophoretic colloids, in
(quasi)-2$d$ configurations, accounting for both dynamically resolved solute-mediated phoretic interactions
and solvent-mediated hydrodynamic interactions. 
Our results show that the system displays different scenarios at changing the colloid-solute affinity 
and it develops a
cluster phase in the chemoattractive case. We study the statistics of cluster 
sizes and cluster morphologies for different magnitudes of colloidal activity. Finally, we provide 
evidences that hydrodynamics plays a relevant role in the aggregation
kinetics and cluster morphology, significantly hindering the cluster growth.
\end{abstract}

\maketitle

\section{Introduction}

Collective behaviour is widespread in Nature: fish schools, insects swarms,
bacterial colonies, plankton blooms are but a few instances of it. 
Collective phenomena in Active Matter are characterized by long-ranged
correlations and large density fluctuations~\cite{MarchettiRMP,RamaARCMP}, 
complex pattern-formation~\cite{Budrene}, and non-equilibrium changes of state, 
such as a flocking~\cite{Vicsek,Cavagna,Ginelli}, clustering~\cite{Peruani2006}, or
mobility induced phase separation~\cite{Cates,Buttinoni}.
Answering key questions on how individuals interact and
communicate goes even beyond fundamental goal of unravelling the physical
mechanisms at the basis of self-organisation in living systems.  
It will help the design of micro- and nano-scale self-propelled 
objects~\cite{Paxton,Dreyfus,Howse,Palacci,Ebbens1,Giomi,BechingerRMP,Ebbens2}, 
with the aim of generating motion in miniaturized devices and 
developping biomimetic environments \cite{Snezhko,Demiroers,GomezSolano,PopescuEPL}. 
Despite most of these natural and artificial particles displace in a fluid medium,
the role played by particle-motion induced hydrodynamic correlations has been essentially overlooked so far. 
Here we present a numerical study of a suspension of non-Brownian colloids which move responding to gradients of a self-generated 
concentration field~\cite{GLA-PRL,GLA-NJP,PopescuEPJEST,Kapral}; the latter determines, dynamically by diffusion and advection, a means of
interaction/communication among the active particles.  
In analogy to typical experimental setups~\cite{Buttinoni,Bocquet,Palacci-Science,Ginot,Maggi},
we consider the  dynamics of a layer of self-propelled colloids (SPCs) on a flat wall  under the
action of gravity embedded in a liquid medium.
We find that the system develops two distinct dynamic regimes, forming large scale
clusters when the interaction of the colloidal particles with solute
is of ``chemoattractive'' type. We characterize the transition between the two
observed non-equilibrium regimes and focus on the morphology and
dynamics of the cluster phase.
With respect to previous studies, we single out quantitatively, for the first time, the
impact of solvent hydrodynamics on the collective dynamics of suspensions of active self-diffusiophoretic Janus colloids.\\

\section{Theory and numerical model}\label{sec:model}

\begin{figure}[htbp]
\begin{center}
  \advance\leftskip-0.55cm
 \includegraphics[scale=0.45]{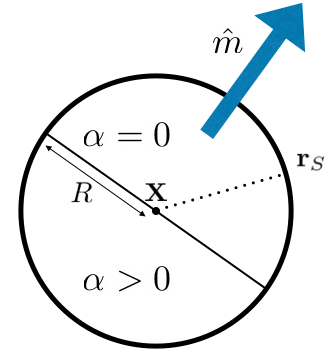}
 \caption{Sketch of a spherical self-phoretic colloid of radius $R$. $\mathbf{X}$ is the position of the centre of mass $\hat{m}$ is the particle characteristic vector,
   based on which we set the activity profile: $\alpha(\mathbf{r}_S)=\alpha_0$ on the bottom hemisphere ($\hat{m} \cdot \hat{n} < 0$) and $\alpha(\mathbf{r}_S) = 0$ on top
 ($\hat{m} \cdot \hat{n}>0$).}
\label{fig0}
\end{center}
\end{figure}

The 3$d$ Navier-Stokes equations
for the fluid (solvent $+$ solute) velocity field $\mathbf{u}$ and for
the solute concentration field $c$,  which read
\begin{eqnarray} \label{eq:eom1}
\partial_t \mathbf{u} + \nabla \cdot (\mathbf{u} \mathbf{u}) &=& -\frac{1}{\rho_f}\nabla P + \nu \nabla^2 \mathbf{u} \\ \nonumber
\partial_t c + \nabla \cdot (\mathbf{u} c) &=& D\nabla^2 c + \mathcal{Q}_c - k_dc,
\end{eqnarray} 
are integrated by means of a hybrid lattice Boltzmann (LB)/finite
difference method~\cite{Succi,Desplat,Stratford,StratfordPago,Swift,Kendon}. Here, $\rho_f$ is the 
fluid density (assumed to be constant, since the flow regime is close to 
incompressible, the maximum Mach number being $Ma \approx 10^{-2}$), $P$ is the pressure
field, $\nu$ and $D$ are, respectively, the kinematic
viscosity and the scalar field diffusivity.
$\mathcal{Q}_c$ represents the production of solute by the
  colloids and it is non-zero only at particle 
surfaces. The local sink term $-k_dc$ models
the degradation of products with rate $k_d$
(with associated characteristic screening length $\ell_d = \sqrt{\frac{D}{k_d}}$ of approximately eight times the particle radius).
The velocities attained in our simulations are such that the typical particle Reynolds $Re = V_p R/\nu$ and P\'eclet $Pe = V_p R/D$ 
numbers are always smaller than $10^{-1}$ ($V_p$ is the self-propulsion speed), thus making the advection terms in 
(\ref{eq:eom1}) negligible.
 The fluid is confined
  along the $z$-direction by two parallel walls (distant about $10$ colloid radii), 
at which a no-slip boundary condition
  is imposed on the velocity field and a zero-flux condition applies
  for the equation for $c$; the system is periodic in $x,y$. 
  Colloids are described as solid spheres of radius $R$, mass $M$ and moment of inertia
    $I = \frac{2}{5}MR^2$, at whose
surfaces momentum and torque exchange between particle and fluid is
implemented via the bounce-back-on-links scheme for LB
probability densities~\cite{Ladd1,Ladd2,Aidun,Nguyen}.
This entails a force $\mathbf{F}_h$ and a torque $\mathbf{T}_h$, exerted by
  the fluid on the particle and resulting from the integrated hydrodynamic stresses
  over the particle surface, that depend on the global configuration of the velocity field $\mathbf{u}$;
  therefore, $\mathbf{F}_h$ and $\mathbf{T}_h$ mediate also, in the general case, hydrodynamic interactions
  among particles. 
  According to the theory of colloidal phoresis~\cite{AndersonARFM}, 
the interaction with a surrounding non-homogeneous concentration field
$c$ induces a flow due to the solute imbalance around the particle surface. This flow, though, is confined to a layer
much thinner than the particle size, since the interaction is typically very short-ranged; consequently, a lubrication
theory analysis leads to account for it as 
an effective boundary condition between the inner layer and the outer fluid,
resulting in the following effective slip velocity for the fluid velocity $\mathbf{u}$ at the particle surface $\Sigma$: 
\begin{equation} \label{eq:effslip}
\mathbf{v}_s = \mu(\mathbf{r}_S) (\mathbf{1} - \hat{r}_S \otimes \hat{r}_S) \cdot \nabla c,
\end{equation}
where the {\it phoretic mobility} $\mu(\mathbf{r}_S)$, $\mathbf{r}_S \in \Sigma$, contains the
details of the colloid-solute interaction. This formulation paves the way, then, to a multi-scale
modelling approach where fluid dynamic processes localised (on molecular scales) close to the particle (and embedded
in the solute-gradient-dependent slip velocity) are effectively
decoupled from large scale flows in the solvent and associated long-range hydrodynamic interactions.
Introducing (\ref{eq:effslip}) in the bounce-back-on-links algorithm~\cite{Ladd1,Ladd2,Aidun,Nguyen} effectively amounts to 
imposing a phoretic force $\mathbf{F}_p$ and torque $\mathbf{T}_p$ on the particle
that take the form:
\begin{eqnarray}
  \mathbf{F}_p &=& -\frac{M}{(4 \pi R^2)\tau_S}\int\int_{\Sigma} \mathbf{v}_s(\mathbf{r}_S)d\mathbf{r}_S \\ \nonumber
  \mathbf{T}_p &=& -\frac{3I}{(8 \pi R^3)\tau_S}\int\int_{\Sigma} \hat{n}(\mathbf{r}_S) \wedge \mathbf{v}_s(\mathbf{r}_S)d\mathbf{r}_S,
\end{eqnarray}
where $\hat{n}$ is the outward normal to the sphere (i.e., with reference to figure \ref{fig0}, is the direction of $\frac{\mathbf{r}_S - \mathbf{X}}{|\mathbf{r}_S - \mathbf{X}|}$)
and $\tau_S$ is the particle Stokes time.
For constant phoretic mobility $\mu(\mathbf{r}_S) \equiv \mu$,
the particle moves with a net propulsion velocity $\mathbf{V}_p \sim -\mu
\nabla c$, directed towards regions of high
concentration of solute for negative $\mu$ (``chemoattractant''), and escaping from it
for positive $\mu$ (``chemorepellent'').
Each particle produces solute $c$ at a constant rate per unit surface, $\alpha_0$, according to the activity profile:
 $$
\alpha(\mathbf{r}_S) =
\left\{
\begin{array}{rl}
\alpha_0 & \mbox{if} \quad \hat{m}\cdot \hat{n} \leq 0 \\
0 & \mbox{if} \quad \hat{m}\cdot \hat{n} > 0,
\end{array}
\right.
  $$
$\hat{m}$ being the particle characteristic unit vector (see the sketch in figure \ref{fig0}); although this particular $\alpha(\mathbf{r}_S)$ specifies {\it Janus}-like
particles~\cite{janus}, in principle the numerical
scheme can deal with arbitrarily {\it patchy}~\cite{Sciortino}  active colloids. 
We stress, furthermore, that the formalism is not specific for self-diffusiophoresis: the only two
required ingredients are, in fact,  the production of a diffusing scalar field and a regime of linear
response of particles to gradients, and as such it enjoys a wider
range of applicability, including systems like 
thermophoretic colloids~\cite{GoleHot,Ripoll1,Ripoll2}, autochemotactic swimmers~\cite{Stark1,Stark2}, gliding
bacteria~\cite{McBride,Peruani2012}.  
The Lagrangian dynamics for the position $\mathbf{X}^{(i)}$ and velocity $\mathbf{V}^{(i)}$
  of the centre of mass
of the $i$-th SPC ($i = 1,2,\dots,N$),  
and for its intrinsic orientation $\hat{m}^{(i)}$ and angular velocity $\Omega^{(i)}$, is described by the 
following equations of motion:
\begin{eqnarray} \label{eq:eom2}
\dot{\mathbf{X}}^{(i)} &=& \mathbf{V}^{(i)} \\ \nonumber
\dot{\mathbf{V}}^{(i)}  &=& \frac{1}{M}\left(\mathbf{F}_h + \mathbf{F}_p
                            + \mathbf{F}_b \right) \\ \nonumber
\dot{\hat{m}}^{(i)}  &=& \Omega^{(i)} \wedge \hat{m}^{(i)}   \\ \nonumber
\dot{\Omega}^{(i)} &=& \frac{1}{I}\left(\mathbf{T}_h + \mathbf{T}_p\right),
\end{eqnarray} 
where $\mathbf{F}_b$ is a generic body force (e.g. gravity) and
the hydrodynmic force $\mathbf{F}_h$ and torque $\mathbf{T}_h$, 
as well as the force $\mathbf{F}_p$ and torque $\mathbf{T}_p$ stemming from the phoretic mechanism, have
been previously introduced.
Equations (\ref{eq:eom2}) are solved by a standard leap-frog algorithm. An isolated free Janus SPC evolving according to (\ref{eq:eom2})
will perform a rectilinear motion with contast speed $V_p = |\mu|\alpha_0/(4D)$~\cite{GLA-PRL,GLA-NJP,PopescuEPJE}.
When assessing the dynamics where no hydrodynamic interactions are present (see section \ref{subsec:nohydro}),
we consider only the equation for $c$ in (\ref{eq:eom1}), with $\mathbf{u} \equiv 0$, whereas $\mathbf{F}_h$ and
$\mathbf{T}_h$ reduce to the usual translational and rotational frictional drag, respectively.
For particles close to contact, lubrication corrections are introduced: the forces and torques acting on two particles 
approaching each other are calculated, in terms of particle velocities and angular velocities, according to a 
grand-resistance-matrix formulation \cite{Nguyen,Janoschek,Brady}.
  In particular, the lubrication correction takes the form of the difference between the lubrication force at a surface
separation $h$ and the force at a given cut-off separation $h_c$; for two particles of radii $R_1$
  and $R_2$ (the particle-wall interaction corresponds to the limit $R_2 \rightarrow \infty$)
  this reads \cite{Nguyen}:
  \begin{equation}\label{eq:lubr}
    \mathbf{F}_{\mbox{\tiny{lub}}}(h) = \left\{
    \begin{array}{ll}
      -6\pi \eta \frac{R_1^2 R_2^2}{(R_1 + R_2)^2}\left(\frac{1}{h} - \frac{1}{h_c}\right)\mathbf{V}_{12}\cdot \hat{r}_{12} & \mbox{if } h \leq h_c \\
      0 & \mbox{if } h > h_c
    \end{array}
    \right.
  \end{equation}
  where $\mathbf{r}_{12} = \mathbf{X}_1 - \mathbf{X}_2 \equiv r_{12} \hat{r}_{12}$ is the particle center-center distance vector and 
  $h = r_{12} - R_1 - R_2$; the cutoff distance is chosen to be $h_c = 0.67$ lattice
  units, which is an optimal value to get good agreement
  with lubrication theory calculations, as shown in \cite{Nguyen}. Lubrication forces may not be
  enough, though, to prevent particle overlap (as recognized also in \cite{Yariv,Varma}), especially when the particle density is large
  (even just locally, as, for instance, inside clusters). Therefore, we add also a short-range
  soft-sphere repulsion modelled by the force $\mathbf{F}_{\mbox{\tiny{ss}}} \propto (h_c^{ss}/h)^3 - 1$,
  with cutoff (coinciding with the soft-sphere radius) $h_c^{ss}=2 h_c$.\\
We have performed numerical simulations of suspensions
with $N=6400$ SPCs on lattices of $1024 \times 1024 \times 24$ grid points
($\approx 410 \, R \times 410 \, R\times 10 \, R$, corresponding to $R=2.5$ lattice spacings, a value
which is relatively small, such to allow simulations of many particle systems, but large enough to keep deviations from the expected
physical behaviour, in terms, e.g., of the drag coefficient, below $10\%$ \cite{Ladd2,Janoschek}),
at fixed area fraction $\phi \approx 0.12$.
Particles are subjected to a gravity force $\mathbf{F}_b$ strong enough to
    prevent them from leaving the bottom wall (the limit fall velocity being five times larger than the self-propulsion speed 
corresponding to the maximum phoretic
    mobility considered, i.e.
  $\frac{F_b}{6 \pi \nu \rho_f R} \approx 5 \frac{|\mu|\alpha}{4D}$). A hard-core
particle-particle repulsion was introduced
to prevent overlapping. Initially particles were randomly
distributed on the surface of the bottom wall. 
Each run lasted approximately $T_{\mbox{\tiny{run}}} \approx 5800\,\tau$.
It is worth noticing that, upon proper non-dimensionalisation by the
  corresponding $\tau$ ($\sim 1\, s$), 
$T_{\mbox{\tiny{run}}}$ is comparable with experimental times \cite{Buttinoni,Bocquet}. 

\section{Results and discussion}

\subsection{Dynamic scenarios controlled by the phoretic mobility.} 
A number of experimental and numerical/theoretical
studies of self-propelled particles in (quasi)-$2d$ 
have given indication of the emergence of 
clustering~\cite{Buttinoni,Bocquet,Palacci-Science,Ginot,Fily,Redner,Pohl1}, however the
nature of the mechanisms determining the formation of aggregates lacks
a consensual agreement and seems to be strongly system-dependent (see also \cite{BechingerRMP} for a recent 
review). 
In our simulations solvent and solute hydrodynamics is fully resolved, from the far field
  down to distances of the order of the particle size (below which it is regularised by the lubrication interaction).
We deal with spherical particles, which rules out 
the possibility of alignment-induced collective motion; instead, chemical production and diffusion 
mediate an effective interaction, 
analogously to the experimental system studied in~\cite{Bocquet,Ginot}.
While in the experiments it was surmised that active colloids felt
an attractive interaction, here we can tune   
the affinity of the particles for the solute via the phoretic
mobility $\mu$, which can be regarded as an effective charge~\cite{GolestanianSoto}, i.e. positive/negative
values induce repulsive/attractive interactions, respectively. 
Indeed, while for $\mu < 0$ our simulations confirm the formation of clusters, for $\mu>0$ such cluster
phase disappears, with the average cluster size going to zero.
Incidentally, let us remark that, in some respect, suspensions of SPCs
  may recall other systems of interacting microswimmers, as, for instance, attractive squirmers \cite{Alarcon}; 
there are however at least two major differences:
 for SPCs, unlike squirmers, the characteristic
 self-propulsion speed is constant only for an isolated swimmer, but in general it depends on the concentration field; 
the second and probably most important one,
 from the point of view of collective dynamics, is that while in the squirmer case particle-particle interactions
 are {\it frozen} (i.e. dictated by the interaction potential once for all), in a SPCs suspension phoretic interactions are mediated 
by the solute field and are, therefore,
  {\it dynamical}, in the sense that they depend on the {\it local} (in space and time) field configuration. 
In other words, phoretic interactions are not pairwise additive
  but change as a function of the global dynamics, and as such, they give rise to a collective behaviour that is genuinely 
out-of-equilibrium.
In what follows, values of $\mu$ will be expressed in units of 
 $\mu_0$, the absolute phoretic mobility for which an isolated particle of radius $R$ would have unitary P\'eclet number.
To address the impact of the chemical affinity on the collective
dynamics quantitatively, we have performed a Voronoj
tessellation analysis of the particle space configurations~\cite{Voronoj,Rycroft}
\footnote[1]{We follow the standard procedure of embedding each particle in a $d$-dimensional cell whose $i$-th
edge (face) is set to be equally distant from the reference particle and its $i$-th nearest neighbour.}.

\begin{figure}[htbp]
\begin{center}
  \advance\leftskip-0.55cm
 \includegraphics[scale=0.45]{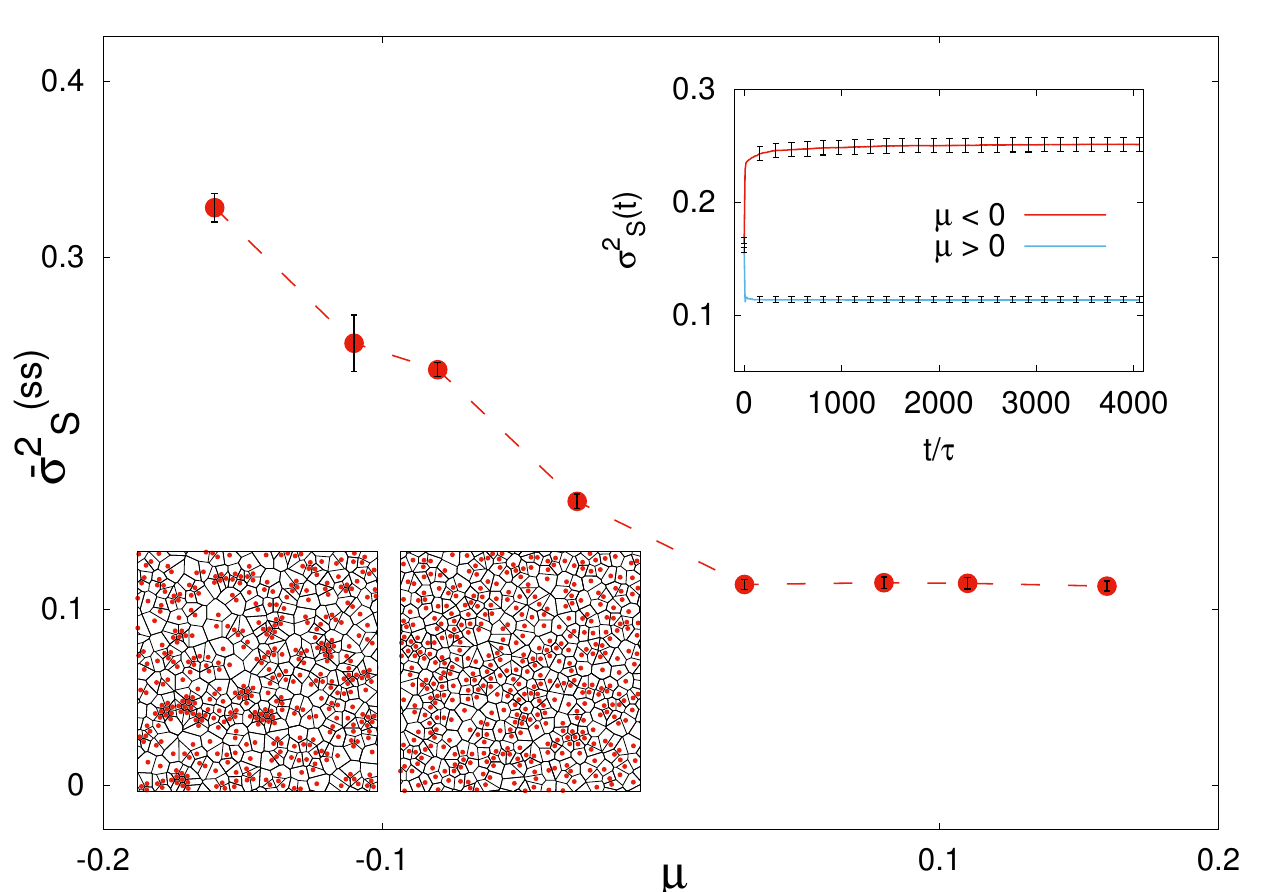}
  \caption{MAIN PANEL: Steady state standard deviation of Voronoi cell
    areas as function
    of the ``coupling constant'' (the phoretic mobility) $\mu$. 
    TOP INSET: $\sigma^2_{\mathcal{S}}(t)$ vs time for two cases with
    positive and negative $\mu$. BOTTOM INSETS: Snapshots of the
    colloid distributions and relative Voronoi diagrams in the
    attractive, $\mu = -0.11$ (LEFT), and repulsive, $\mu = +0.11$ (RIGHT), case,
    respectively.}
\label{fig1}
\end{center}
\end{figure}

The bottom insets of Fig.\ref{fig1} show the Voronoi diagrams both for
the repulsive and cluster-forming regimes; as clearly visible to the naked
eye, the geometries of the Voronoi cells for chemoattractive and chemorepellent
active colloids are distinctively different. 
The standard deviation of the cell area distribution
$\sigma_{\mathcal{S}}^2 (t) \equiv \frac{1}{N\overline{\mathcal{S}}^2}
\sum_{i=1}^N (\mathcal{S}_i -
 \overline{\mathcal{S}})^2$
(normalized by the square of the mean value $\overline{\mathcal{S}}$)
turns out to be a good indicator to distinguish the two types of dynamics.
In the top inset of Fig.~\ref{fig1} we plot $\sigma_{\mathcal{S}}(t)$,
as function of time (given in units of $\tau = R/V_p$, i.e. the time an isolated particle 
takes to displace by one radius), for two cases with different 
sign of the phoretic mobility. In the attractive case, $\mu < 0$, cluster
formation induces the appearance of very small (and large) cells 
and, hence, the surface fluctuations grow and eventually saturate at long times. 
For positive $\mu$, instead, colloids repel each other and tend to reach an optimal covering of
the space, implying that  $\sigma^2_{\mathcal{S}}(t)$  attains a (lower) value
which remains constant in time. 
Correspondingly, the dependence of $\overline{\sigma^2}_{\mathcal{S}}^{(SS)}$ (the time average of
$\overline{\sigma^2}_{\mathcal{S}}(t)$ over the steady state) on $\mu$ discriminates between the two regimes:
it is high for negatively large $\mu$, decreases as $\mu$ approaches zero and then stays low and constant for $\mu>0$.\\
We will focus, in what follows, on the chemoattractive case, but before moving on we stress that 
the phase diagram for chemorepulsive self-phoretic colloids, as
recently shown theoretically and numerically \cite{Benno,Pohl2}, is indeed rather complex
and deserves further investigation. 

\subsection{Cluster statistics and morphology.}  
\begin{figure}[htbp]
\begin{center}
  \advance\leftskip-0.55cm
  \includegraphics[scale=0.6]{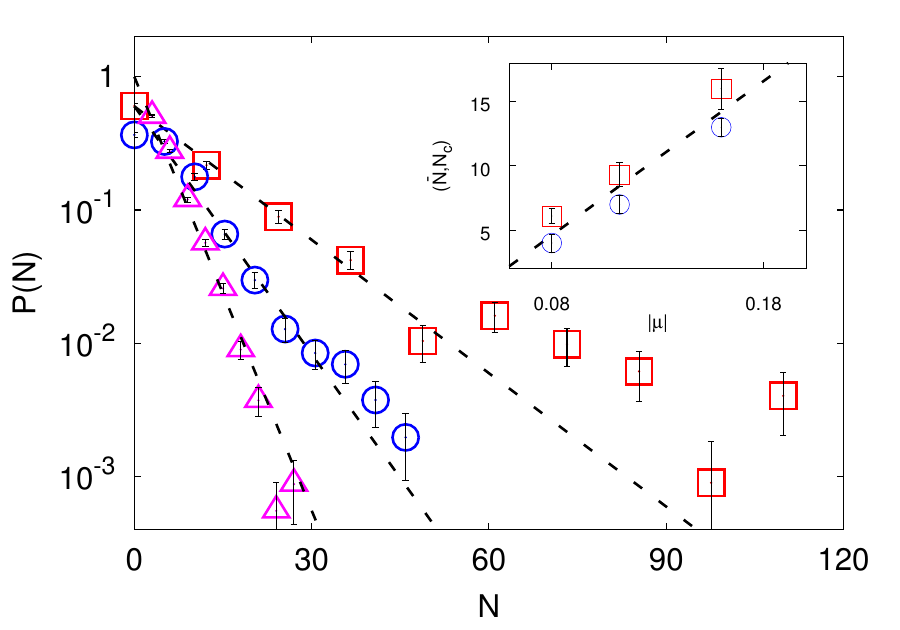}
  \caption{MAIN PANEL: PDFs of cluster sizes for three values of the phoretic
    mobility: $\mu = -0.16$ (\textcolor{red}{$\Box$}), $\mu = -0.11$
    (\textcolor{blue}{$\bigcirc$}) and $\mu = -0.08$
    (\textcolor{magenta}{$\triangle$}); 
    the dashed lines represent exponential fit which are drawn to
    guide the eye. INSET: Characteristic ($n_c$, \textcolor{blue}{$\bigcirc$}) and mean
    ($\overline{n}$, \textcolor{red}{$\Box$}) cluster sizes as function of
    $|\mu| \propto V_p$, the intrisic SPC velocity (the dashed line indicates a linear relation).}
\label{fig2}
\end{center}
\end{figure}

\begin{figure}[htbp]
\begin{center}
  \advance\leftskip-0.55cm
  \includegraphics[scale=0.6]{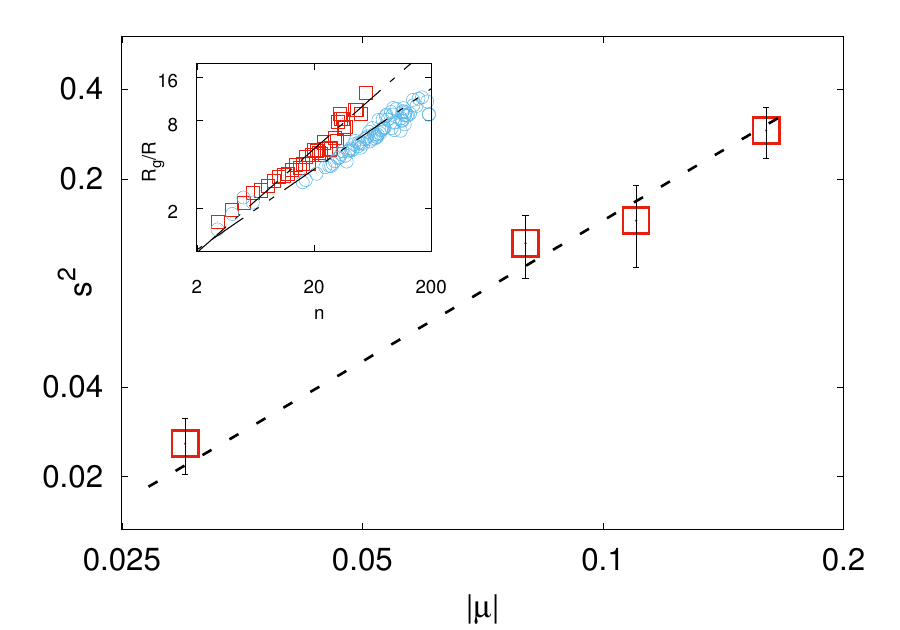}
  \caption{MAIN PANEL: Deviation of the steady state SPC density
    fluctuations $\sigma^2_{\rho}$ from the value for inactive
    particles $\sigma^2_{\rho_0}$, normalized as $s^2 = \frac{\sigma^2_{\rho} - \sigma^2_{\rho_0}}{\sigma^2_{\rho_0} - \langle \rho \rangle^2}$, vs the
    coupling strength $|\mu|$ from LB (squares) and the
    phenomenological derivation (\ref{eq:last}), with fractal
    dimension $d_f = 1.4$, as measured in  the simulations. INSET: Mean gyration radius
    of clusters vs number of particles with (\textcolor{red}{$\Box$}) and without
    (\textcolor{blue}{$\bigcirc$}) HI. The two dashed lines represent the power
    law $R_g \sim n^{1/d_f}$ ($d_f = 1.8$ for no-HI).}
\label{fig3}
\end{center}
\end{figure}

We first characterize the 
cluster size distribution of chemoattractive SPCs at changing
the colloid/solute coupling intensity, $|\mu|$. We identify clusters
according to a distance criterion. 
Two particles share a  {\it bond} if their centers are at a distance equal to or less than a
cutoff apart \footnote[3]{We set such cutoff to the value of  $\Lambda = 2R + h$, $h$ being
the hard-core range of interaction.}, and we define as clusters, groups of particles
connected to each other through a bond. We compute probability density functions
(PDFs) of cluster sizes over the steady state of each run. Fig.~\ref{fig2} shows such PDFs, which can be in all cases fitted 
to an exponential, $\mathcal{P}(n) \propto e^{-n/n_c}$, over a wide
range of sizes $n$.  The characteristic value $n_c$ and the mean size
$\overline{n} = \frac{1}{N_{clus}}\sum_{i=1}^{N_{clus}}n_i$ increase linearly with $|\mu|$ (see inset of
figure \ref{fig2}), hence with the velocity of an isolated particle, in
agreement with experimental and numerical observations~\cite{Buttinoni,Bocquet,Pohl1}. 
The global attractor for the SPCs dynamics is a set $\mathcal{S} = \bigcup_{i=1}^{N_{clus}} \mathcal{C}_i$,
where $\mathcal{C}_i$ is the $i$-th cluster with surface of area
$\mathcal{A}_i$ and containing $n_i$ SPCs. 
Correspondingly, the colloid number density becomes  $\rho(x) = \rho_i
= n_i/\mathcal{A}_i$ if $ x \in \mathcal{C}_i$ and zero otherwise. 
Since the colloid density fluctuations can be expressed as
$\sigma^2_{\rho} = \langle (\rho(x) - \langle \rho \rangle)^2 \rangle$
(where $\langle \dots \rangle$ denotes a surface average), we arrive at
\begin{equation} \label{eq:sigmavsmu}
\sigma_{\rho}^2 \propto \langle \rho(x)^2 \rangle = \sum_{i=1}^{N_{clus}}  \frac{\rho_i^2\mathcal{A}_i}{|\Sigma|}
 \mathcal{P}(\mathcal{A}_i),
\end{equation}
where $|\Sigma|$ is the measure of the whole plane, and $\mathcal{P}(\mathcal{A}_i)$ is the probability of having 
a cluster of area $\mathcal{A}_i$. The number of particles in a cluster $n$ is known to scale with the
cluster gyration radius $R_g$ as $n \sim R_g^{d_f}$, $d_f$ being the
fractal (Hausdorff) dimension~\cite{Witten,MeakinPRL}; hence, the density of
the $i$-th cluster, 
$\rho_i$, will  behave as $\rho_i \sim
\mathcal{A}_i^{d_f/2-1}$. The exponential behavior of $\mathcal{P}(n)$
predicts that 
\begin{equation} \label{eq:last}
\sigma_{\rho}^2\sim
 \sigma_{\rho_0}^2  (1+a|\mu|)^{\zeta(d_f)} , \;\ \quad \zeta(d_f) = \frac{3d_f-2}{d_f},
\end{equation}
where $\sigma_{\rho_0}^2$ stands for  the
fluctuations for a inactive particles and $a$ is a phenomenological
parameter, and where we have used the relation
$n_c \sim |\mu|$ \footnote[4]{In principle there can be a dependence on $\mu$ also of the
fractal dimension $d_f$ ; we assume here, however, that the change in $\mu$  affects only the characteristic cluster size and
not its ``compactness'' (or fractality).} (see inset of figure
\ref{fig2}). Fig.~\ref{fig3} shows the quantitative agreement of the
predicted power law, with the correct scaling exponent, with the numerical observations.

\subsection{Role of hydrodynamic interactions.}\label{subsec:nohydro}  
\begin{figure}
\begin{center}
  \advance\leftskip-0.55cm
  \includegraphics[scale=0.6]{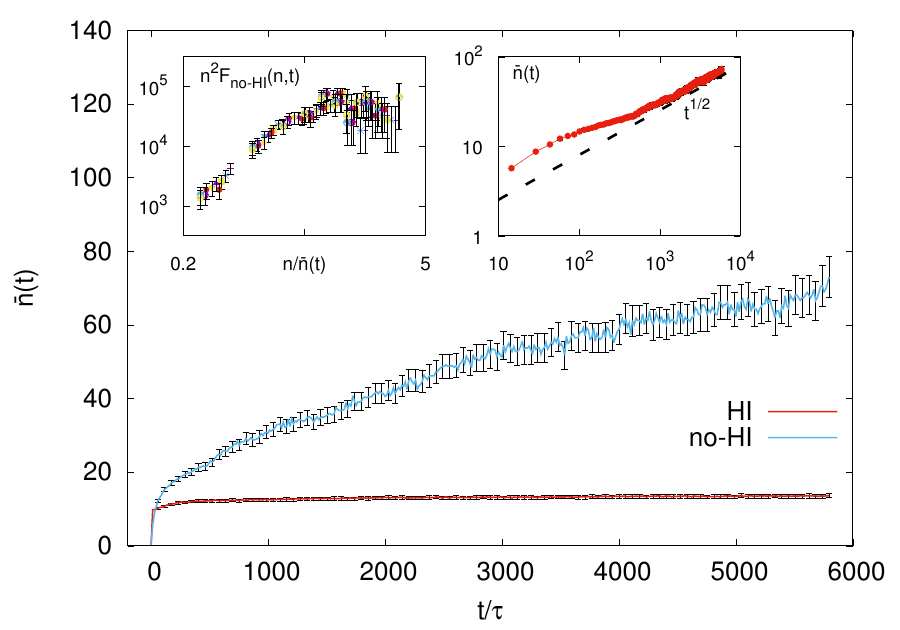}
  \caption{MAIN PANEL: Mean cluster size $\overline{n}$ vs time from
    simulations with (HI) and without (no-HI) hydrodynamic
    interactions (the values for the HI case are magnified by a factor two, for the sake of clarity of visualisation).
    LEFT INSET: Cluster size distributions for $\mu = -0.16$
    and no-HI, at various times $t \in [1500\tau; 3000\tau]$ during the coarsening process.  
   RIGHT INSET: Log-log plot of $\overline{n}$ vs $t$, without HI, highlighting the scaling $t^{1/2}$ in the coarsening process.}
\label{fig4}
\end{center}
\end{figure}

\begin{figure}
\begin{center}
  \advance\leftskip-0.55cm
  \includegraphics[scale=0.6]{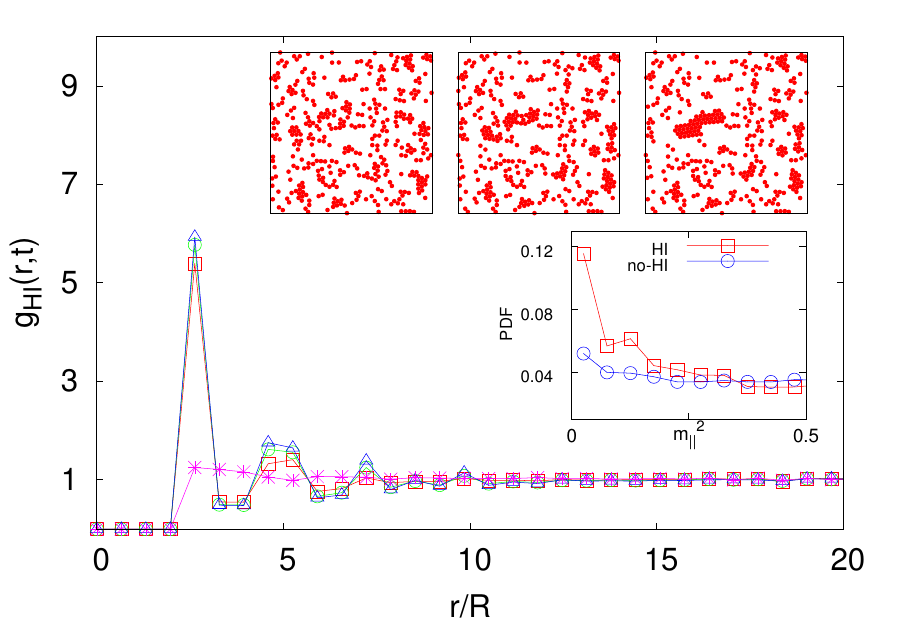}
  \includegraphics[scale=0.6]{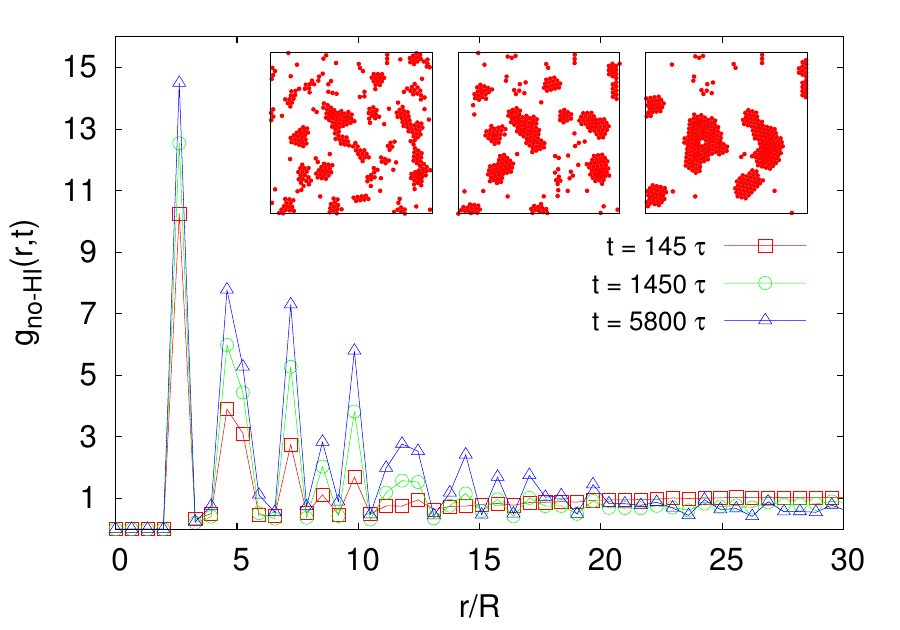}
  \caption{RDFs for $\mu=-0.16$ at three
    different times and corresponding snapshots of the colloid
    distribution, indicating cluster formation, in a sub-system of size
    $256 \times 256$, located in the center of the box. The two panels
    correspond to simulations with (TOP) and without (BOTTOM)
    hydrodynamic interactions among colloids. In the top figure
    also the repulsive case $\mu=+0.16$ ($\textcolor{magenta}{\ast}$) is
    reported for comparison. 
   TOP PANEL INSET: PDFs of $m_{||}^2$,
    the square magnitude projection onto the $\{x,y\}$-plane of the
    colloid orientations, from simulations with (HI) and without (no-HI)
    hydrodynamics.}
\label{fig5}
\end{center}
\end{figure}

Self-propelled colloids
interact due to both the chemicals they produce and the
flows they induce. Understanding the relative magnitude
and competition between these two sources of dynamic interactions
remains challenging. As described in section \ref{sec:model}, the model put forward allows us to switch off
the hydrodynamic interactions (HI), yet keeping the self-phoretic mechanism and the correct translational and rotational hydrodynamic friction.
Interestingly, our study reveals that, although the different dynamic scenarios at changing the sign of
the phoretic mobility are preserved even without HI (being mainly determined 
by the chemical interaction), HI have a profound effect in the
kinetics of formation and morphology of the observed aggregates.
In the absence of particle induced flows in the solvent, attractive
SPCs ($\mu<0$) show an enhanced tendency to form
clusters, as it appears in figure Fig.~\ref{fig4}, where we compare the time evolution of the mean cluster size
$\overline{n}(t)$, with and without HI (no-HI). 
In the no-HI case, clusters coarsen, with $\overline{n}(t)$ growing in time as $t^{1/2}$ (top right inset), as for
  domains in a spinodal decomposition. 
  The same behaviour ($\overline{n}(t) \sim t^{1/2}$) has been observed, indeed,
    in simulations of self-propelled Brownian particles
  interacting via a shifted-truncated Lennard-Jones potential \cite{Redner,Alarcon}. 
  With HI, instead, coarsening is arrested, as observed in experiments \cite{Bocquet}.
  Simulations have suggested that in suspensions of attractive squirmers
    the emergence of continuous or arrested coarsening is selected
    depending on the form and intensity of the active stress (the coefficient $B_2$ in the
    squirmer terminology) \cite{Alarcon};
    self-phoretic Janus colloids behave, in this respect, as squirmers
    with $B_2=0$ \cite{PopescuNEW}, for which, indeed, arrested coarsening was observed \cite{Alarcon}.
  Due to the non stationarity of the coarsening process, steady state PDFs of cluster sizes 
  cannot be computed in the no-HI simulations. Nevertheless, we observe that istantaneous 
  cluster size distributions $F(n,t)$ (i.e. the number of clusters of size $n$ at time $t$) 
  tend to assume a self-preserving scaling form 
  $F(n,t) \sim n^{-2}f(n/\overline{n}(t))$, as it happens in classical colloidal aggregation phenomena
  for mass-conserving systems \cite{MeakinRevGeo}.
  This is shown in the top left inset of Fig.~\ref{fig4}, where we plot $n^2F(n,t)$ vs
  $n/\overline{n}(t)$ and see that all sets of points for different
  $t$'s in the coarsening regime where 
  $\overline{n} \sim t^{1/2}$, within error bars, collapse onto each other.
The statement on the different dynamics, with and without HI, is corroborated by the inspection of the radial
distribution functions (RDFs)~\cite{Hansen} (indicated as $g_{HI}(r,t)$
and $g_{no-HI}(r,t)$, respectively), defined as the probability of finding a particle between the distances $r$ and $r+dr$ from a reference
  particle (and averaged over all particles), i.e.
  $$
 g(r,t) = \frac{1}{\rho_N N} \sum_{i=1}^N\sum_{\substack{j=1 \\ j \neq i}}^N \delta\left(\mathbf{r}-\mathbf{X}_j(t) + \mathbf{X}_i(t) \right),
  $$
where $\rho_N$ is the particle number density and $\delta(x)$ is the Dirac's delta.
RDFs at different times are shown in Fig.~\ref{fig5}: 
without HI the peaks are higher and decay more slowly, associated to the development
of clusters larger than those formed when hydrodynamics is active.
Besides, clusters appear substantially more compact, as appreciated in
the snapshots (insets) and quantified by the measurement of a larger
fractal dimension ($d_f^{(HI)} \approx 1.4$ and $d_f^{(no-HI)} \approx 1.8$, see figure \ref{fig3}). 
Hydrodynamics hinders, then, the colloidal aggregation process. Several complex
mechanisms can be conjectured to cause this phenomenon: dynamically
induced effective repulsion among particles, fluid flow generated
disturbances in the chemical field distribution, etc. 
An effect, that we could clearly identify, is
an enhanced tendency of SPCs to be oriented {\it off-plane}, 
when HI are present, not only when they hit a cluster (as, e.g., in the mechanism proposed
in \cite{ZoettlStark}) but also for isolated particles.
This may be attributed to fluid motion close to the wall, giving rise to
hydrodynamic torques that rotate the particles.
Actually the roto-translational dynamics of self-diffusiophoretic colloids, at and close to interfaces, is
an intricate problem \cite{Das,Uspal,Mozaffari}: 
in highly confined situations, one might indeed expect even the opposite trend (clusterisation enhancement)
\cite{ZoettlStark}; the argument maintains, therefore, a qualitative character.
Nevertheless, the inset in the Fig.~\ref{fig5} (top panel) provides a quantitative insight to the picture.
There, we show the PDF of the degree of alignment of the 
particle orientation with the bounding solid wall, $m_{||}^2 = m_x^2+m_y^2$. When HI are present, indeed,
the peak of the PDF around $m_{||}^2 \sim 0$ is more pronounced,
{\sl i.e.} there is a larger fraction of colloids pointing
out of the plane. Accordingly, the self-propulsion speed is effectively reduced, 
thus limiting the {\it in-plane} mobility and diminisihing the capability of particles to gather and clusterize.
Before concluding, we would like to remind that it is still an open issue whether Janus particles can really have a
homogeneous phoretic mobility; if the opposite is true, i.e. inhomogeneous $\mu(\mathbf{r})$ gives rise,
in response to gradients of the concentration field, to {\it chemical} torques that can
lead, themselves, to clustering inhibition \cite{Saha,Bickel,Liebchen2017}.
With hydrodynamic interactions, the dynamics is, of course, even more complicated, due to the competition of
these effects, and it is subject of ongoing research.

\section{Conclusions}
To conclude, we have used a mesoscopic numerical model of fully resolved spherical active colloids,
propelled by self-generated gradients of a scalar field (e.g. a
chemical product) where the self-induced hydrodynamic flows can be accounted for.
We have identified the role of the phoretic mobility as the key controlling parameter 
that determines two distinct 
dynamic regimes and the onset of a cluster phase. By means
of a Voronoi tessellation we have characterized the cluster state
finding that the probability distribution of sizes decays
exponentially with a mean size growing linearly with the
particle activity, in agreement with experimental
results~\cite{Bocquet,Buttinoni}. 
We have quantified the profound effect that hydrodynamics plays
inhibiting clustering for negative phoretic 
mobilities.
We have identified the interplay between induced flows and
particle reorientation as a possible explanation to the strong slowing down of
cluster coarsening, although it remains an open question, which needs a deeper analysis, whether fluid-wall interactions dominate
  over particle-particle hydrodynamic correlations.
This study shows that our novel
numerical method is powerful and enjoys some unique features, namely
the explicit description of chemical signalling, through the
production and diffusion of a solute concentration field and the
solvent hydrodynamics, to simulate realistic systems.
Moreover, it opens the way to address the dynamics of self-propelled
colloids in general geometries and also for stronger activity (larger $Pe$), both in isotropic and unforced situations, where
aggregation can lead to the formation of active colloidal gels, or
under gravity as in the experimental sedimentation setup.

\section*{Acknowledgements}
We acknowledge the Spanish MINECO and Generalitat de Catalunya DURSI
for financial support under the projects FIS2015-67837-P and 2014SGR-922, respectively. 
I.P.  acknowledges  {\sl Generalitat de Catalunya  } under Program
{\sl Icrea Acad\`emia}. 
This work was possible thanks to the access to MareNostrum
Supercomputer at Barcelona Supercomputing Center (BSC) and also
through the Partnership for Advanced Computing in Europe (PRACE).

\bibliography{spc} 

\begin{thebibliography}{78}%
\makeatletter
\providecommand \@ifxundefined [1]{%
 \@ifx{#1\undefined}
}%
\providecommand \@ifnum [1]{%
 \ifnum #1\expandafter \@firstoftwo
 \else \expandafter \@secondoftwo
 \fi
}%
\providecommand \@ifx [1]{%
 \ifx #1\expandafter \@firstoftwo
 \else \expandafter \@secondoftwo
 \fi
}%
\providecommand \natexlab [1]{#1}%
\providecommand \enquote  [1]{``#1''}%
\providecommand \bibnamefont  [1]{#1}%
\providecommand \bibfnamefont [1]{#1}%
\providecommand \citenamefont [1]{#1}%
\providecommand \href@noop [0]{\@secondoftwo}%
\providecommand \href [0]{\begingroup \@sanitize@url \@href}%
\providecommand \@href[1]{\@@startlink{#1}\@@href}%
\providecommand \@@href[1]{\endgroup#1\@@endlink}%
\providecommand \@sanitize@url [0]{\catcode `\\12\catcode `\$12\catcode
  `\&12\catcode `\#12\catcode `\^12\catcode `\_12\catcode `\%12\relax}%
\providecommand \@@startlink[1]{}%
\providecommand \@@endlink[0]{}%
\providecommand \url  [0]{\begingroup\@sanitize@url \@url }%
\providecommand \@url [1]{\endgroup\@href {#1}{\urlprefix }}%
\providecommand \urlprefix  [0]{URL }%
\providecommand \Eprint [0]{\href }%
\providecommand \doibase [0]{http://dx.doi.org/}%
\providecommand \selectlanguage [0]{\@gobble}%
\providecommand \bibinfo  [0]{\@secondoftwo}%
\providecommand \bibfield  [0]{\@secondoftwo}%
\providecommand \translation [1]{[#1]}%
\providecommand \BibitemOpen [0]{}%
\providecommand \bibitemStop [0]{}%
\providecommand \bibitemNoStop [0]{.\EOS\space}%
\providecommand \EOS [0]{\spacefactor3000\relax}%
\providecommand \BibitemShut  [1]{\csname bibitem#1\endcsname}%
\let\auto@bib@innerbib\@empty
\bibitem [{\citenamefont {Marchetti}\ \emph {et~al.}(2013)\citenamefont
  {Marchetti}, \citenamefont {Joanny}, \citenamefont {Ramaswamy}, \citenamefont
  {Liverpool}, \citenamefont {Prost}, \citenamefont {Rao},\ and\ \citenamefont
  {Simha}}]{MarchettiRMP}%
  \BibitemOpen
  \bibfield  {author} {\bibinfo {author} {\bibfnamefont {M.}~\bibnamefont
  {Marchetti}}, \bibinfo {author} {\bibfnamefont {J.-F.}\ \bibnamefont
  {Joanny}}, \bibinfo {author} {\bibfnamefont {S.}~\bibnamefont {Ramaswamy}},
  \bibinfo {author} {\bibfnamefont {T.}~\bibnamefont {Liverpool}}, \bibinfo
  {author} {\bibfnamefont {J.}~\bibnamefont {Prost}}, \bibinfo {author}
  {\bibfnamefont {M.}~\bibnamefont {Rao}}, \ and\ \bibinfo {author}
  {\bibfnamefont {R.}~\bibnamefont {Simha}},\ }\href@noop {} {\bibfield
  {journal} {\bibinfo  {journal} {Rev. Mod. Phys.}\ }\textbf {\bibinfo {volume}
  {85}},\ \bibinfo {pages} {1143} (\bibinfo {year} {2013})}\BibitemShut
  {NoStop}%
\bibitem [{\citenamefont {Ramaswamy}(2010)}]{RamaARCMP}%
  \BibitemOpen
  \bibfield  {author} {\bibinfo {author} {\bibfnamefont {S.}~\bibnamefont
  {Ramaswamy}},\ }\href@noop {} {\bibfield  {journal} {\bibinfo  {journal}
  {Annu. Rev. Cond. Matter Phys}\ }\textbf {\bibinfo {volume} {1}},\ \bibinfo
  {pages} {323} (\bibinfo {year} {2010})}\BibitemShut {NoStop}%
\bibitem [{\citenamefont {Budrene}\ and\ \citenamefont {Berg}(1991)}]{Budrene}%
  \BibitemOpen
  \bibfield  {author} {\bibinfo {author} {\bibfnamefont {E.}~\bibnamefont
  {Budrene}}\ and\ \bibinfo {author} {\bibfnamefont {H.}~\bibnamefont {Berg}},\
  }\href@noop {} {\bibfield  {journal} {\bibinfo  {journal} {Nature}\ }\textbf
  {\bibinfo {volume} {349}},\ \bibinfo {pages} {630} (\bibinfo {year}
  {1991})}\BibitemShut {NoStop}%
\bibitem [{\citenamefont {Vicsek}\ \emph {et~al.}(1995)\citenamefont {Vicsek},
  \citenamefont {Czir\'ok}, \citenamefont {Ben-Jacob}, \citenamefont {Cohen},\
  and\ \citenamefont {Shochet}}]{Vicsek}%
  \BibitemOpen
  \bibfield  {author} {\bibinfo {author} {\bibfnamefont {T.}~\bibnamefont
  {Vicsek}}, \bibinfo {author} {\bibfnamefont {T.}~\bibnamefont {Czir\'ok}},
  \bibinfo {author} {\bibfnamefont {E.}~\bibnamefont {Ben-Jacob}}, \bibinfo
  {author} {\bibfnamefont {I.}~\bibnamefont {Cohen}}, \ and\ \bibinfo {author}
  {\bibfnamefont {O.}~\bibnamefont {Shochet}},\ }\href@noop {} {\bibfield
  {journal} {\bibinfo  {journal} {Phys. Rev. Lett.}\ }\textbf {\bibinfo
  {volume} {75}},\ \bibinfo {pages} {1226} (\bibinfo {year}
  {1995})}\BibitemShut {NoStop}%
\bibitem [{\citenamefont {Cavagna}\ and\ \citenamefont
  {Giardina}(2014)}]{Cavagna}%
  \BibitemOpen
  \bibfield  {author} {\bibinfo {author} {\bibfnamefont {A.}~\bibnamefont
  {Cavagna}}\ and\ \bibinfo {author} {\bibfnamefont {I.}~\bibnamefont
  {Giardina}},\ }\href@noop {} {\bibfield  {journal} {\bibinfo  {journal}
  {Annu. Rec. Cond. Matter Phys.}\ }\textbf {\bibinfo {volume} {5}},\ \bibinfo
  {pages} {183} (\bibinfo {year} {2014})}\BibitemShut {NoStop}%
\bibitem [{\citenamefont {Ginelli}\ and\ \citenamefont
  {Chat\'e}(2010)}]{Ginelli}%
  \BibitemOpen
  \bibfield  {author} {\bibinfo {author} {\bibfnamefont {F.}~\bibnamefont
  {Ginelli}}\ and\ \bibinfo {author} {\bibfnamefont {H.}~\bibnamefont
  {Chat\'e}},\ }\href@noop {} {\bibfield  {journal} {\bibinfo  {journal} {Phys.
  Rev. Lett.}\ }\textbf {\bibinfo {volume} {105}},\ \bibinfo {pages} {1681032}
  (\bibinfo {year} {2010})}\BibitemShut {NoStop}%
\bibitem [{\citenamefont {Peruani}\ \emph {et~al.}(2006)\citenamefont
  {Peruani}, \citenamefont {Deutsch},\ and\ \citenamefont
  {B\"ar}}]{Peruani2006}%
  \BibitemOpen
  \bibfield  {author} {\bibinfo {author} {\bibfnamefont {F.}~\bibnamefont
  {Peruani}}, \bibinfo {author} {\bibfnamefont {A.}~\bibnamefont {Deutsch}}, \
  and\ \bibinfo {author} {\bibfnamefont {M.}~\bibnamefont {B\"ar}},\
  }\href@noop {} {\bibfield  {journal} {\bibinfo  {journal} {Phys. Rev. E}\
  }\textbf {\bibinfo {volume} {74}},\ \bibinfo {pages} {030904} (\bibinfo
  {year} {2006})}\BibitemShut {NoStop}%
\bibitem [{\citenamefont {Cates}\ \emph {et~al.}(2010)\citenamefont {Cates},
  \citenamefont {Marenduzzo}, \citenamefont {Pagonabarrga},\ and\ \citenamefont
  {Tailleur}}]{Cates}%
  \BibitemOpen
  \bibfield  {author} {\bibinfo {author} {\bibfnamefont {M.}~\bibnamefont
  {Cates}}, \bibinfo {author} {\bibfnamefont {D.}~\bibnamefont {Marenduzzo}},
  \bibinfo {author} {\bibfnamefont {I.}~\bibnamefont {Pagonabarrga}}, \ and\
  \bibinfo {author} {\bibfnamefont {J.}~\bibnamefont {Tailleur}},\ }\href@noop
  {} {\bibfield  {journal} {\bibinfo  {journal} {Proc. Natl. Acad. Sci. USA}\
  }\textbf {\bibinfo {volume} {107}},\ \bibinfo {pages} {11715} (\bibinfo
  {year} {2010})}\BibitemShut {NoStop}%
\bibitem [{\citenamefont {Buttinoni}\ \emph {et~al.}(2013)\citenamefont
  {Buttinoni}, \citenamefont {Bialk\'e}, \citenamefont {K\"ummel},
  \citenamefont {L\"owen}, \citenamefont {Bechinger},\ and\ \citenamefont
  {Speck}}]{Buttinoni}%
  \BibitemOpen
  \bibfield  {author} {\bibinfo {author} {\bibfnamefont {I.}~\bibnamefont
  {Buttinoni}}, \bibinfo {author} {\bibfnamefont {J.}~\bibnamefont {Bialk\'e}},
  \bibinfo {author} {\bibfnamefont {F.}~\bibnamefont {K\"ummel}}, \bibinfo
  {author} {\bibfnamefont {H.}~\bibnamefont {L\"owen}}, \bibinfo {author}
  {\bibfnamefont {C.}~\bibnamefont {Bechinger}}, \ and\ \bibinfo {author}
  {\bibfnamefont {T.}~\bibnamefont {Speck}},\ }\href@noop {} {\bibfield
  {journal} {\bibinfo  {journal} {Phys. Rev. Lett.}\ }\textbf {\bibinfo
  {volume} {110}},\ \bibinfo {pages} {238301} (\bibinfo {year}
  {2013})}\BibitemShut {NoStop}%
\bibitem [{\citenamefont {Paxton}\ \emph {et~al.}(2004)\citenamefont {Paxton},
  \citenamefont {Kistler}, \citenamefont {Olmeda}, \citenamefont {Sen},
  \citenamefont {Angelo}, \citenamefont {Cao}, \citenamefont {Mallouk},
  \citenamefont {Lammert},\ and\ \citenamefont {Crespi}}]{Paxton}%
  \BibitemOpen
  \bibfield  {author} {\bibinfo {author} {\bibfnamefont {W.}~\bibnamefont
  {Paxton}}, \bibinfo {author} {\bibfnamefont {K.}~\bibnamefont {Kistler}},
  \bibinfo {author} {\bibfnamefont {C.}~\bibnamefont {Olmeda}}, \bibinfo
  {author} {\bibfnamefont {A.}~\bibnamefont {Sen}}, \bibinfo {author}
  {\bibfnamefont {S.~K.~S.}\ \bibnamefont {Angelo}}, \bibinfo {author}
  {\bibfnamefont {Y.}~\bibnamefont {Cao}}, \bibinfo {author} {\bibfnamefont
  {T.}~\bibnamefont {Mallouk}}, \bibinfo {author} {\bibfnamefont
  {P.}~\bibnamefont {Lammert}}, \ and\ \bibinfo {author} {\bibfnamefont
  {V.}~\bibnamefont {Crespi}},\ }\href@noop {} {\bibfield  {journal} {\bibinfo
  {journal} {J. Am. Chem. Soc.}\ }\textbf {\bibinfo {volume} {126}},\ \bibinfo
  {pages} {13424} (\bibinfo {year} {2004})}\BibitemShut {NoStop}%
\bibitem [{\citenamefont {Dreyfus}\ \emph {et~al.}(2005)\citenamefont
  {Dreyfus}, \citenamefont {Baudry}, \citenamefont {Roper}, \citenamefont
  {Fermigier}, \citenamefont {Stone},\ and\ \citenamefont {Bibette}}]{Dreyfus}%
  \BibitemOpen
  \bibfield  {author} {\bibinfo {author} {\bibfnamefont {R.}~\bibnamefont
  {Dreyfus}}, \bibinfo {author} {\bibfnamefont {J.}~\bibnamefont {Baudry}},
  \bibinfo {author} {\bibfnamefont {M.}~\bibnamefont {Roper}}, \bibinfo
  {author} {\bibfnamefont {M.}~\bibnamefont {Fermigier}}, \bibinfo {author}
  {\bibfnamefont {H.}~\bibnamefont {Stone}}, \ and\ \bibinfo {author}
  {\bibfnamefont {J.}~\bibnamefont {Bibette}},\ }\href@noop {} {\bibfield
  {journal} {\bibinfo  {journal} {Nature}\ }\textbf {\bibinfo {volume} {437}},\
  \bibinfo {pages} {862} (\bibinfo {year} {2005})}\BibitemShut {NoStop}%
\bibitem [{\citenamefont {Howse}\ \emph {et~al.}(2007)\citenamefont {Howse},
  \citenamefont {Jones}, \citenamefont {Ryan}, \citenamefont {Gough},
  \citenamefont {Vafabakhsh},\ and\ \citenamefont {Golestanian}}]{Howse}%
  \BibitemOpen
  \bibfield  {author} {\bibinfo {author} {\bibfnamefont {J.}~\bibnamefont
  {Howse}}, \bibinfo {author} {\bibfnamefont {R.}~\bibnamefont {Jones}},
  \bibinfo {author} {\bibfnamefont {A.}~\bibnamefont {Ryan}}, \bibinfo {author}
  {\bibfnamefont {T.}~\bibnamefont {Gough}}, \bibinfo {author} {\bibfnamefont
  {R.}~\bibnamefont {Vafabakhsh}}, \ and\ \bibinfo {author} {\bibfnamefont
  {R.}~\bibnamefont {Golestanian}},\ }\href@noop {} {\bibfield  {journal}
  {\bibinfo  {journal} {Phys. Rev. Lett.}\ }\textbf {\bibinfo {volume} {99}},\
  \bibinfo {pages} {048102} (\bibinfo {year} {2007})}\BibitemShut {NoStop}%
\bibitem [{\citenamefont {Palacci}\ \emph {et~al.}(2010)\citenamefont
  {Palacci}, \citenamefont {Cottin-Bizonne}, \citenamefont {Ybert},\ and\
  \citenamefont {Bocquet}}]{Palacci}%
  \BibitemOpen
  \bibfield  {author} {\bibinfo {author} {\bibfnamefont {J.}~\bibnamefont
  {Palacci}}, \bibinfo {author} {\bibfnamefont {C.}~\bibnamefont
  {Cottin-Bizonne}}, \bibinfo {author} {\bibfnamefont {C.}~\bibnamefont
  {Ybert}}, \ and\ \bibinfo {author} {\bibfnamefont {L.}~\bibnamefont
  {Bocquet}},\ }\href@noop {} {\bibfield  {journal} {\bibinfo  {journal} {Phys.
  Rev. Lett.}\ }\textbf {\bibinfo {volume} {105}},\ \bibinfo {pages} {088304}
  (\bibinfo {year} {2010})}\BibitemShut {NoStop}%
\bibitem [{\citenamefont {Ebbens}\ and\ \citenamefont {Howse}(2010)}]{Ebbens1}%
  \BibitemOpen
  \bibfield  {author} {\bibinfo {author} {\bibfnamefont {S.}~\bibnamefont
  {Ebbens}}\ and\ \bibinfo {author} {\bibfnamefont {J.}~\bibnamefont {Howse}},\
  }\href@noop {} {\bibfield  {journal} {\bibinfo  {journal} {Soft Matter}\
  }\textbf {\bibinfo {volume} {6}},\ \bibinfo {pages} {726} (\bibinfo {year}
  {2010})}\BibitemShut {NoStop}%
\bibitem [{\citenamefont {Giomi}\ \emph {et~al.}(2013)\citenamefont {Giomi},
  \citenamefont {Hawley-Weld},\ and\ \citenamefont {Mahadevan}}]{Giomi}%
  \BibitemOpen
  \bibfield  {author} {\bibinfo {author} {\bibfnamefont {L.}~\bibnamefont
  {Giomi}}, \bibinfo {author} {\bibfnamefont {N.}~\bibnamefont {Hawley-Weld}},
  \ and\ \bibinfo {author} {\bibfnamefont {L.}~\bibnamefont {Mahadevan}},\
  }\href@noop {} {\bibfield  {journal} {\bibinfo  {journal} {Proc. R. Soc. A}\
  }\textbf {\bibinfo {volume} {469}},\ \bibinfo {pages} {20120637} (\bibinfo
  {year} {2013})}\BibitemShut {NoStop}%
\bibitem [{\citenamefont {Bechinger}\ \emph {et~al.}(2016)\citenamefont
  {Bechinger}, \citenamefont {Leonardo}, \citenamefont {L\"owen}, \citenamefont
  {Reichhardt}, \citenamefont {Volpe},\ and\ \citenamefont
  {Volpe}}]{BechingerRMP}%
  \BibitemOpen
  \bibfield  {author} {\bibinfo {author} {\bibfnamefont {C.}~\bibnamefont
  {Bechinger}}, \bibinfo {author} {\bibfnamefont {R.~D.}\ \bibnamefont
  {Leonardo}}, \bibinfo {author} {\bibfnamefont {H.}~\bibnamefont {L\"owen}},
  \bibinfo {author} {\bibfnamefont {C.}~\bibnamefont {Reichhardt}}, \bibinfo
  {author} {\bibfnamefont {G.}~\bibnamefont {Volpe}}, \ and\ \bibinfo {author}
  {\bibfnamefont {G.}~\bibnamefont {Volpe}},\ }\href@noop {} {\bibfield
  {journal} {\bibinfo  {journal} {Rev. Mod. Phys.}\ }\textbf {\bibinfo {volume}
  {88}},\ \bibinfo {pages} {045006} (\bibinfo {year} {2016})}\BibitemShut
  {NoStop}%
\bibitem [{\citenamefont {Ebbens}(2016)}]{Ebbens2}%
  \BibitemOpen
  \bibfield  {author} {\bibinfo {author} {\bibfnamefont {S.}~\bibnamefont
  {Ebbens}},\ }\href@noop {} {\bibfield  {journal} {\bibinfo  {journal} {Curr.
  Opin. Colloid Interface Sci.}\ }\textbf {\bibinfo {volume} {21}},\ \bibinfo
  {pages} {14} (\bibinfo {year} {2016})}\BibitemShut {NoStop}%
\bibitem [{\citenamefont {Snezhko}\ and\ \citenamefont
  {Aranson}(2011)}]{Snezhko}%
  \BibitemOpen
  \bibfield  {author} {\bibinfo {author} {\bibfnamefont {A.}~\bibnamefont
  {Snezhko}}\ and\ \bibinfo {author} {\bibfnamefont {I.}~\bibnamefont
  {Aranson}},\ }\href@noop {} {\bibfield  {journal} {\bibinfo  {journal} {Nat.
  Mater.}\ }\textbf {\bibinfo {volume} {10}},\ \bibinfo {pages} {698} (\bibinfo
  {year} {2011})}\BibitemShut {NoStop}%
\bibitem [{\citenamefont {Demir\"ors}\ \emph {et~al.}(2017)\citenamefont
  {Demir\"ors}, \citenamefont {Eichenseher}, \citenamefont {Loessner},\ and\
  \citenamefont {Studart}}]{Demiroers}%
  \BibitemOpen
  \bibfield  {author} {\bibinfo {author} {\bibfnamefont {A.}~\bibnamefont
  {Demir\"ors}}, \bibinfo {author} {\bibfnamefont {F.}~\bibnamefont
  {Eichenseher}}, \bibinfo {author} {\bibfnamefont {M.}~\bibnamefont
  {Loessner}}, \ and\ \bibinfo {author} {\bibfnamefont {A.}~\bibnamefont
  {Studart}},\ }\href@noop {} {\bibfield  {journal} {\bibinfo  {journal} {Nat.
  Comm.}\ }\textbf {\bibinfo {volume} {8}},\ \bibinfo {pages} {1872} (\bibinfo
  {year} {2017})}\BibitemShut {NoStop}%
\bibitem [{\citenamefont {G\'omez-Solano}\ \emph {et~al.}(2017)\citenamefont
  {G\'omez-Solano}, \citenamefont {Samin}, \citenamefont {Lozano},
  \citenamefont {Ruedas-Batuecas}, \citenamefont {van Roij},\ and\
  \citenamefont {Bechinger}}]{GomezSolano}%
  \BibitemOpen
  \bibfield  {author} {\bibinfo {author} {\bibfnamefont {J.}~\bibnamefont
  {G\'omez-Solano}}, \bibinfo {author} {\bibfnamefont {S.}~\bibnamefont
  {Samin}}, \bibinfo {author} {\bibfnamefont {C.}~\bibnamefont {Lozano}},
  \bibinfo {author} {\bibfnamefont {P.}~\bibnamefont {Ruedas-Batuecas}},
  \bibinfo {author} {\bibfnamefont {R.}~\bibnamefont {van Roij}}, \ and\
  \bibinfo {author} {\bibfnamefont {C.}~\bibnamefont {Bechinger}},\ }\href@noop
  {} {\bibfield  {journal} {\bibinfo  {journal} {Sci. Rep.}\ }\textbf {\bibinfo
  {volume} {7}},\ \bibinfo {pages} {14891} (\bibinfo {year}
  {2017})}\BibitemShut {NoStop}%
\bibitem [{\citenamefont {Popescu}\ \emph {et~al.}(2011)\citenamefont
  {Popescu}, \citenamefont {Tasinkevych},\ and\ \citenamefont
  {Dietrich}}]{PopescuEPL}%
  \BibitemOpen
  \bibfield  {author} {\bibinfo {author} {\bibfnamefont {M.}~\bibnamefont
  {Popescu}}, \bibinfo {author} {\bibfnamefont {M.}~\bibnamefont
  {Tasinkevych}}, \ and\ \bibinfo {author} {\bibfnamefont {S.}~\bibnamefont
  {Dietrich}},\ }\href@noop {} {\bibfield  {journal} {\bibinfo  {journal}
  {Europhys. Lett.}\ }\textbf {\bibinfo {volume} {95}},\ \bibinfo {pages}
  {28004} (\bibinfo {year} {2011})}\BibitemShut {NoStop}%
\bibitem [{\citenamefont {Golestanian}\ \emph {et~al.}(2005)\citenamefont
  {Golestanian}, \citenamefont {Liverpool},\ and\ \citenamefont
  {Ajdari}}]{GLA-PRL}%
  \BibitemOpen
  \bibfield  {author} {\bibinfo {author} {\bibfnamefont {R.}~\bibnamefont
  {Golestanian}}, \bibinfo {author} {\bibfnamefont {T.}~\bibnamefont
  {Liverpool}}, \ and\ \bibinfo {author} {\bibfnamefont {A.}~\bibnamefont
  {Ajdari}},\ }\href@noop {} {\bibfield  {journal} {\bibinfo  {journal} {Phys.
  Rev. Lett.}\ }\textbf {\bibinfo {volume} {94}},\ \bibinfo {pages} {220801}
  (\bibinfo {year} {2005})}\BibitemShut {NoStop}%
\bibitem [{\citenamefont {Golestanian}\ \emph {et~al.}(2007)\citenamefont
  {Golestanian}, \citenamefont {Liverpool},\ and\ \citenamefont
  {Ajdari}}]{GLA-NJP}%
  \BibitemOpen
  \bibfield  {author} {\bibinfo {author} {\bibfnamefont {R.}~\bibnamefont
  {Golestanian}}, \bibinfo {author} {\bibfnamefont {T.}~\bibnamefont
  {Liverpool}}, \ and\ \bibinfo {author} {\bibfnamefont {A.}~\bibnamefont
  {Ajdari}},\ }\href@noop {} {\bibfield  {journal} {\bibinfo  {journal} {New J.
  Phys.}\ }\textbf {\bibinfo {volume} {9}},\ \bibinfo {pages} {126} (\bibinfo
  {year} {2007})}\BibitemShut {NoStop}%
\bibitem [{\citenamefont {Popescu}\ \emph {et~al.}(2016)\citenamefont
  {Popescu}, \citenamefont {Uspal},\ and\ \citenamefont
  {Dietrich}}]{PopescuEPJEST}%
  \BibitemOpen
  \bibfield  {author} {\bibinfo {author} {\bibfnamefont {M.}~\bibnamefont
  {Popescu}}, \bibinfo {author} {\bibfnamefont {W.}~\bibnamefont {Uspal}}, \
  and\ \bibinfo {author} {\bibfnamefont {S.}~\bibnamefont {Dietrich}},\
  }\href@noop {} {\bibfield  {journal} {\bibinfo  {journal} {Eur. Phys. J.
  Special Topics}\ }\textbf {\bibinfo {volume} {225}},\ \bibinfo {pages} {2189}
  (\bibinfo {year} {2016})}\BibitemShut {NoStop}%
\bibitem [{\citenamefont {R\"uckner}\ and\ \citenamefont
  {Kapral}(2007)}]{Kapral}%
  \BibitemOpen
  \bibfield  {author} {\bibinfo {author} {\bibfnamefont {G.}~\bibnamefont
  {R\"uckner}}\ and\ \bibinfo {author} {\bibfnamefont {R.}~\bibnamefont
  {Kapral}},\ }\href@noop {} {\bibfield  {journal} {\bibinfo  {journal} {Phys.
  Rev. Lett.}\ }\textbf {\bibinfo {volume} {98}},\ \bibinfo {pages} {150603}
  (\bibinfo {year} {2007})}\BibitemShut {NoStop}%
\bibitem [{\citenamefont {Theurkauff}\ \emph {et~al.}(2012)\citenamefont
  {Theurkauff}, \citenamefont {Cottin-Bizonne}, \citenamefont {Palacci},
  \citenamefont {Ybert},\ and\ \citenamefont {Bocquet}}]{Bocquet}%
  \BibitemOpen
  \bibfield  {author} {\bibinfo {author} {\bibfnamefont {I.}~\bibnamefont
  {Theurkauff}}, \bibinfo {author} {\bibfnamefont {C.}~\bibnamefont
  {Cottin-Bizonne}}, \bibinfo {author} {\bibfnamefont {J.}~\bibnamefont
  {Palacci}}, \bibinfo {author} {\bibfnamefont {C.}~\bibnamefont {Ybert}}, \
  and\ \bibinfo {author} {\bibfnamefont {L.}~\bibnamefont {Bocquet}},\
  }\href@noop {} {\bibfield  {journal} {\bibinfo  {journal} {Phys. Rev. Lett.}\
  }\textbf {\bibinfo {volume} {108}},\ \bibinfo {pages} {268303} (\bibinfo
  {year} {2012})}\BibitemShut {NoStop}%
\bibitem [{\citenamefont {Palacci}\ \emph {et~al.}(2013)\citenamefont
  {Palacci}, \citenamefont {Sacanna}, \citenamefont {Steinberg}, \citenamefont
  {Pine},\ and\ \citenamefont {Chaikin}}]{Palacci-Science}%
  \BibitemOpen
  \bibfield  {author} {\bibinfo {author} {\bibfnamefont {J.}~\bibnamefont
  {Palacci}}, \bibinfo {author} {\bibfnamefont {S.}~\bibnamefont {Sacanna}},
  \bibinfo {author} {\bibfnamefont {A.}~\bibnamefont {Steinberg}}, \bibinfo
  {author} {\bibfnamefont {D.}~\bibnamefont {Pine}}, \ and\ \bibinfo {author}
  {\bibfnamefont {P.}~\bibnamefont {Chaikin}},\ }\href@noop {} {\bibfield
  {journal} {\bibinfo  {journal} {Science}\ }\textbf {\bibinfo {volume}
  {339}},\ \bibinfo {pages} {936} (\bibinfo {year} {2013})}\BibitemShut
  {NoStop}%
\bibitem [{\citenamefont {Ginot}\ \emph {et~al.}(2018)\citenamefont {Ginot},
  \citenamefont {Theurkauff}, \citenamefont {Detcheverry}, \citenamefont
  {Ybert},\ and\ \citenamefont {Cottin-Bizonne}}]{Ginot}%
  \BibitemOpen
  \bibfield  {author} {\bibinfo {author} {\bibfnamefont {F.}~\bibnamefont
  {Ginot}}, \bibinfo {author} {\bibfnamefont {I.}~\bibnamefont {Theurkauff}},
  \bibinfo {author} {\bibfnamefont {F.}~\bibnamefont {Detcheverry}}, \bibinfo
  {author} {\bibfnamefont {C.}~\bibnamefont {Ybert}}, \ and\ \bibinfo {author}
  {\bibfnamefont {C.}~\bibnamefont {Cottin-Bizonne}},\ }\href@noop {}
  {\bibfield  {journal} {\bibinfo  {journal} {Nat. Comm.}\ }\textbf {\bibinfo
  {volume} {9}},\ \bibinfo {pages} {696} (\bibinfo {year} {2018})}\BibitemShut
  {NoStop}%
\bibitem [{\citenamefont {Maggi}\ \emph {et~al.}(2015)\citenamefont {Maggi},
  \citenamefont {Simmechen}, \citenamefont {Saglimbeni}, \citenamefont
  {Dipalo}, \citenamefont {Angelis}, \citenamefont {S\'anchez},\ and\
  \citenamefont {Leonardo}}]{Maggi}%
  \BibitemOpen
  \bibfield  {author} {\bibinfo {author} {\bibfnamefont {C.}~\bibnamefont
  {Maggi}}, \bibinfo {author} {\bibfnamefont {J.}~\bibnamefont {Simmechen}},
  \bibinfo {author} {\bibfnamefont {F.}~\bibnamefont {Saglimbeni}}, \bibinfo
  {author} {\bibfnamefont {M.}~\bibnamefont {Dipalo}}, \bibinfo {author}
  {\bibfnamefont {F.~D.}\ \bibnamefont {Angelis}}, \bibinfo {author}
  {\bibfnamefont {S.}~\bibnamefont {S\'anchez}}, \ and\ \bibinfo {author}
  {\bibfnamefont {R.~D.}\ \bibnamefont {Leonardo}},\ }\href@noop {} {\bibfield
  {journal} {\bibinfo  {journal} {Small}\ }\textbf {\bibinfo {volume} {12}},\
  \bibinfo {pages} {446} (\bibinfo {year} {2015})}\BibitemShut {NoStop}%
\bibitem [{\citenamefont {Succi}(2018)}]{Succi}%
  \BibitemOpen
  \bibfield  {author} {\bibinfo {author} {\bibfnamefont {S.}~\bibnamefont
  {Succi}},\ }\href@noop {} {\emph {\bibinfo {title} {The lattice Boltzmann
  equation for complex states of flowing matter}}}\ (\bibinfo  {publisher}
  {Oxford University Press},\ \bibinfo {year} {2018})\BibitemShut {NoStop}%
\bibitem [{\citenamefont {Desplat}\ \emph {et~al.}(2001)\citenamefont
  {Desplat}, \citenamefont {Pagonabarraga},\ and\ \citenamefont
  {Bladon}}]{Desplat}%
  \BibitemOpen
  \bibfield  {author} {\bibinfo {author} {\bibfnamefont {J.-C.}\ \bibnamefont
  {Desplat}}, \bibinfo {author} {\bibfnamefont {I.}~\bibnamefont
  {Pagonabarraga}}, \ and\ \bibinfo {author} {\bibfnamefont {P.}~\bibnamefont
  {Bladon}},\ }\href@noop {} {\bibfield  {journal} {\bibinfo  {journal} {Comp.
  Phys. Comm.}\ }\textbf {\bibinfo {volume} {134}},\ \bibinfo {pages} {273}
  (\bibinfo {year} {2001})}\BibitemShut {NoStop}%
\bibitem [{\citenamefont {Stratford}\ \emph {et~al.}(2005)\citenamefont
  {Stratford}, \citenamefont {Adhikari}, \citenamefont {Pagonabarraga},\ and\
  \citenamefont {Desplat}}]{Stratford}%
  \BibitemOpen
  \bibfield  {author} {\bibinfo {author} {\bibfnamefont {K.}~\bibnamefont
  {Stratford}}, \bibinfo {author} {\bibfnamefont {R.}~\bibnamefont {Adhikari}},
  \bibinfo {author} {\bibfnamefont {I.}~\bibnamefont {Pagonabarraga}}, \ and\
  \bibinfo {author} {\bibfnamefont {J.-C.}\ \bibnamefont {Desplat}},\
  }\href@noop {} {\bibfield  {journal} {\bibinfo  {journal} {J. Stat. Phys.}\
  }\textbf {\bibinfo {volume} {121}},\ \bibinfo {pages} {163} (\bibinfo {year}
  {2005})}\BibitemShut {NoStop}%
\bibitem [{\citenamefont {Stratford}\ and\ \citenamefont
  {Pagonabarraga}(2008)}]{StratfordPago}%
  \BibitemOpen
  \bibfield  {author} {\bibinfo {author} {\bibfnamefont {K.}~\bibnamefont
  {Stratford}}\ and\ \bibinfo {author} {\bibfnamefont {I.}~\bibnamefont
  {Pagonabarraga}},\ }\href@noop {} {\bibfield  {journal} {\bibinfo  {journal}
  {Comput. Math. Appl.}\ }\textbf {\bibinfo {volume} {55}},\ \bibinfo {pages}
  {1585} (\bibinfo {year} {2008})}\BibitemShut {NoStop}%
\bibitem [{\citenamefont {Swift}\ \emph {et~al.}(1996)\citenamefont {Swift},
  \citenamefont {Orlandini}, \citenamefont {Osborn},\ and\ \citenamefont
  {Yeomans}}]{Swift}%
  \BibitemOpen
  \bibfield  {author} {\bibinfo {author} {\bibfnamefont {M.}~\bibnamefont
  {Swift}}, \bibinfo {author} {\bibfnamefont {E.}~\bibnamefont {Orlandini}},
  \bibinfo {author} {\bibfnamefont {W.}~\bibnamefont {Osborn}}, \ and\ \bibinfo
  {author} {\bibfnamefont {J.}~\bibnamefont {Yeomans}},\ }\href@noop {}
  {\bibfield  {journal} {\bibinfo  {journal} {Phys. Rev. E}\ }\textbf {\bibinfo
  {volume} {54}},\ \bibinfo {pages} {5041} (\bibinfo {year}
  {1996})}\BibitemShut {NoStop}%
\bibitem [{\citenamefont {Kendon}\ \emph {et~al.}(2001)\citenamefont {Kendon},
  \citenamefont {Cates}, \citenamefont {Pagonabarraga}, \citenamefont
  {Desplat},\ and\ \citenamefont {Bladon}}]{Kendon}%
  \BibitemOpen
  \bibfield  {author} {\bibinfo {author} {\bibfnamefont {V.}~\bibnamefont
  {Kendon}}, \bibinfo {author} {\bibfnamefont {M.}~\bibnamefont {Cates}},
  \bibinfo {author} {\bibfnamefont {I.}~\bibnamefont {Pagonabarraga}}, \bibinfo
  {author} {\bibfnamefont {J.-C.}\ \bibnamefont {Desplat}}, \ and\ \bibinfo
  {author} {\bibfnamefont {P.}~\bibnamefont {Bladon}},\ }\href@noop {}
  {\bibfield  {journal} {\bibinfo  {journal} {J. Fluid Mech.}\ }\textbf
  {\bibinfo {volume} {440}},\ \bibinfo {pages} {147} (\bibinfo {year}
  {2001})}\BibitemShut {NoStop}%
\bibitem [{\citenamefont {Ladd}(1994{\natexlab{a}})}]{Ladd1}%
  \BibitemOpen
  \bibfield  {author} {\bibinfo {author} {\bibfnamefont {A.}~\bibnamefont
  {Ladd}},\ }\href@noop {} {\bibfield  {journal} {\bibinfo  {journal} {J. Fluid
  Mech.}\ }\textbf {\bibinfo {volume} {271}},\ \bibinfo {pages} {285} (\bibinfo
  {year} {1994}{\natexlab{a}})}\BibitemShut {NoStop}%
\bibitem [{\citenamefont {Ladd}(1994{\natexlab{b}})}]{Ladd2}%
  \BibitemOpen
  \bibfield  {author} {\bibinfo {author} {\bibfnamefont {A.}~\bibnamefont
  {Ladd}},\ }\href@noop {} {\bibfield  {journal} {\bibinfo  {journal} {J. Fluid
  Mech.}\ }\textbf {\bibinfo {volume} {271}},\ \bibinfo {pages} {311} (\bibinfo
  {year} {1994}{\natexlab{b}})}\BibitemShut {NoStop}%
\bibitem [{\citenamefont {Aidun}\ \emph {et~al.}(1998)\citenamefont {Aidun},
  \citenamefont {Lu},\ and\ \citenamefont {Ding}}]{Aidun}%
  \BibitemOpen
  \bibfield  {author} {\bibinfo {author} {\bibfnamefont {C.}~\bibnamefont
  {Aidun}}, \bibinfo {author} {\bibfnamefont {Y.}~\bibnamefont {Lu}}, \ and\
  \bibinfo {author} {\bibfnamefont {E.-J.}\ \bibnamefont {Ding}},\ }\href@noop
  {} {\bibfield  {journal} {\bibinfo  {journal} {J. Fluid Mech.}\ }\textbf
  {\bibinfo {volume} {373}},\ \bibinfo {pages} {287} (\bibinfo {year}
  {1998})}\BibitemShut {NoStop}%
\bibitem [{\citenamefont {Nguyen}\ and\ \citenamefont {Ladd}(2002)}]{Nguyen}%
  \BibitemOpen
  \bibfield  {author} {\bibinfo {author} {\bibfnamefont {N.-Q.}\ \bibnamefont
  {Nguyen}}\ and\ \bibinfo {author} {\bibfnamefont {A.}~\bibnamefont {Ladd}},\
  }\href@noop {} {\bibfield  {journal} {\bibinfo  {journal} {Phys. Rev. E}\
  }\textbf {\bibinfo {volume} {66}},\ \bibinfo {pages} {046708} (\bibinfo
  {year} {2002})}\BibitemShut {NoStop}%
\bibitem [{\citenamefont {Anderson}(1989)}]{AndersonARFM}%
  \BibitemOpen
  \bibfield  {author} {\bibinfo {author} {\bibfnamefont {J.}~\bibnamefont
  {Anderson}},\ }\href@noop {} {\bibfield  {journal} {\bibinfo  {journal}
  {Annu. Rev. Fluid Mech.}\ }\textbf {\bibinfo {volume} {21}},\ \bibinfo
  {pages} {61} (\bibinfo {year} {1989})}\BibitemShut {NoStop}%
\bibitem [{\citenamefont {Walther}\ and\ \citenamefont
  {M\"uller}(2008)}]{janus}%
  \BibitemOpen
  \bibfield  {author} {\bibinfo {author} {\bibfnamefont {A.}~\bibnamefont
  {Walther}}\ and\ \bibinfo {author} {\bibfnamefont {A.}~\bibnamefont
  {M\"uller}},\ }\href@noop {} {\bibfield  {journal} {\bibinfo  {journal} {Soft
  Matter}\ }\textbf {\bibinfo {volume} {4}},\ \bibinfo {pages} {663} (\bibinfo
  {year} {2008})}\BibitemShut {NoStop}%
\bibitem [{\citenamefont {Sciortino}\ \emph {et~al.}(2009)\citenamefont
  {Sciortino}, \citenamefont {Giacometti},\ and\ \citenamefont
  {Pastore}}]{Sciortino}%
  \BibitemOpen
  \bibfield  {author} {\bibinfo {author} {\bibfnamefont {F.}~\bibnamefont
  {Sciortino}}, \bibinfo {author} {\bibfnamefont {A.}~\bibnamefont
  {Giacometti}}, \ and\ \bibinfo {author} {\bibfnamefont {G.}~\bibnamefont
  {Pastore}},\ }\href@noop {} {\bibfield  {journal} {\bibinfo  {journal} {Phys.
  Rev. Lett.}\ }\textbf {\bibinfo {volume} {103}},\ \bibinfo {pages} {237801}
  (\bibinfo {year} {2009})}\BibitemShut {NoStop}%
\bibitem [{\citenamefont {Golestanian}(2012)}]{GoleHot}%
  \BibitemOpen
  \bibfield  {author} {\bibinfo {author} {\bibfnamefont {R.}~\bibnamefont
  {Golestanian}},\ }\href@noop {} {\bibfield  {journal} {\bibinfo  {journal}
  {Phys. Rev. Lett.}\ }\textbf {\bibinfo {volume} {108}},\ \bibinfo {pages}
  {038303} (\bibinfo {year} {2012})}\BibitemShut {NoStop}%
\bibitem [{\citenamefont {L\"usebrink}\ and\ \citenamefont
  {Ripoll}(2012)}]{Ripoll1}%
  \BibitemOpen
  \bibfield  {author} {\bibinfo {author} {\bibfnamefont {D.}~\bibnamefont
  {L\"usebrink}}\ and\ \bibinfo {author} {\bibfnamefont {M.}~\bibnamefont
  {Ripoll}},\ }\href@noop {} {\bibfield  {journal} {\bibinfo  {journal} {J.
  Chem. Phys.}\ }\textbf {\bibinfo {volume} {136}},\ \bibinfo {pages} {084106}
  (\bibinfo {year} {2012})}\BibitemShut {NoStop}%
\bibitem [{\citenamefont {Yang}\ \emph {et~al.}(2014)\citenamefont {Yang},
  \citenamefont {Wysocki},\ and\ \citenamefont {Ripoll}}]{Ripoll2}%
  \BibitemOpen
  \bibfield  {author} {\bibinfo {author} {\bibfnamefont {M.}~\bibnamefont
  {Yang}}, \bibinfo {author} {\bibfnamefont {A.}~\bibnamefont {Wysocki}}, \
  and\ \bibinfo {author} {\bibfnamefont {M.}~\bibnamefont {Ripoll}},\
  }\href@noop {} {\bibfield  {journal} {\bibinfo  {journal} {Soft Matter}\
  }\textbf {\bibinfo {volume} {10}},\ \bibinfo {pages} {6208} (\bibinfo {year}
  {2014})}\BibitemShut {NoStop}%
\bibitem [{\citenamefont {Taktikos}\ \emph {et~al.}(2011)\citenamefont
  {Taktikos}, \citenamefont {Zaburdaev},\ and\ \citenamefont {Stark}}]{Stark1}%
  \BibitemOpen
  \bibfield  {author} {\bibinfo {author} {\bibfnamefont {J.}~\bibnamefont
  {Taktikos}}, \bibinfo {author} {\bibfnamefont {V.}~\bibnamefont {Zaburdaev}},
  \ and\ \bibinfo {author} {\bibfnamefont {H.}~\bibnamefont {Stark}},\
  }\href@noop {} {\bibfield  {journal} {\bibinfo  {journal} {Phys. Rev. E}\
  }\textbf {\bibinfo {volume} {84}},\ \bibinfo {pages} {041924} (\bibinfo
  {year} {2011})}\BibitemShut {NoStop}%
\bibitem [{\citenamefont {Taktikos}\ \emph {et~al.}(2012)\citenamefont
  {Taktikos}, \citenamefont {Zaburdaev},\ and\ \citenamefont {Stark}}]{Stark2}%
  \BibitemOpen
  \bibfield  {author} {\bibinfo {author} {\bibfnamefont {J.}~\bibnamefont
  {Taktikos}}, \bibinfo {author} {\bibfnamefont {V.}~\bibnamefont {Zaburdaev}},
  \ and\ \bibinfo {author} {\bibfnamefont {H.}~\bibnamefont {Stark}},\
  }\href@noop {} {\bibfield  {journal} {\bibinfo  {journal} {Phys. Rev. E}\
  }\textbf {\bibinfo {volume} {85}},\ \bibinfo {pages} {051901} (\bibinfo
  {year} {2012})}\BibitemShut {NoStop}%
\bibitem [{\citenamefont {McBride}(2001)}]{McBride}%
  \BibitemOpen
  \bibfield  {author} {\bibinfo {author} {\bibfnamefont {M.}~\bibnamefont
  {McBride}},\ }\href@noop {} {\bibfield  {journal} {\bibinfo  {journal} {Annu.
  Rev. Microbiol.}\ }\textbf {\bibinfo {volume} {55}},\ \bibinfo {pages} {49}
  (\bibinfo {year} {2001})}\BibitemShut {NoStop}%
\bibitem [{\citenamefont {Peruani}\ \emph {et~al.}(2012)\citenamefont
  {Peruani}, \citenamefont {Starru{\ss}}, \citenamefont {S{\o}gaard-Andersen},
  \citenamefont {Deutsch},\ and\ \citenamefont {B\"ar}}]{Peruani2012}%
  \BibitemOpen
  \bibfield  {author} {\bibinfo {author} {\bibfnamefont {F.}~\bibnamefont
  {Peruani}}, \bibinfo {author} {\bibfnamefont {J.}~\bibnamefont
  {Starru{\ss}}}, \bibinfo {author} {\bibfnamefont {L.}~\bibnamefont
  {S{\o}gaard-Andersen}}, \bibinfo {author} {\bibfnamefont {A.}~\bibnamefont
  {Deutsch}}, \ and\ \bibinfo {author} {\bibfnamefont {M.}~\bibnamefont
  {B\"ar}},\ }\href@noop {} {\bibfield  {journal} {\bibinfo  {journal} {Phys.
  Rev. Lett.}\ }\textbf {\bibinfo {volume} {108}},\ \bibinfo {pages} {098102}
  (\bibinfo {year} {2012})}\BibitemShut {NoStop}%
\bibitem [{\citenamefont {Popescu}\ \emph {et~al.}(2010)\citenamefont
  {Popescu}, \citenamefont {Dietrich}, \citenamefont {Tasinkevych},\ and\
  \citenamefont {Ralston}}]{PopescuEPJE}%
  \BibitemOpen
  \bibfield  {author} {\bibinfo {author} {\bibfnamefont {M.}~\bibnamefont
  {Popescu}}, \bibinfo {author} {\bibfnamefont {S.}~\bibnamefont {Dietrich}},
  \bibinfo {author} {\bibfnamefont {M.}~\bibnamefont {Tasinkevych}}, \ and\
  \bibinfo {author} {\bibfnamefont {J.}~\bibnamefont {Ralston}},\ }\href@noop
  {} {\bibfield  {journal} {\bibinfo  {journal} {Eur. Phys. J. E}\ }\textbf
  {\bibinfo {volume} {31}},\ \bibinfo {pages} {351} (\bibinfo {year}
  {2010})}\BibitemShut {NoStop}%
\bibitem [{\citenamefont {Janoschek}\ \emph {et~al.}(2013)\citenamefont
  {Janoschek}, \citenamefont {Harting},\ and\ \citenamefont
  {Toschi}}]{Janoschek}%
  \BibitemOpen
  \bibfield  {author} {\bibinfo {author} {\bibfnamefont {F.}~\bibnamefont
  {Janoschek}}, \bibinfo {author} {\bibfnamefont {J.}~\bibnamefont {Harting}},
  \ and\ \bibinfo {author} {\bibfnamefont {F.}~\bibnamefont {Toschi}},\
  }\href@noop {} {\bibfield  {journal} {\bibinfo  {journal} {arXiv:1308.6482}\
  } (\bibinfo {year} {2013})}\BibitemShut {NoStop}%
\bibitem [{\citenamefont {Brady}\ and\ \citenamefont {Bossis}(1988)}]{Brady}%
  \BibitemOpen
  \bibfield  {author} {\bibinfo {author} {\bibfnamefont {J.}~\bibnamefont
  {Brady}}\ and\ \bibinfo {author} {\bibfnamefont {G.}~\bibnamefont {Bossis}},\
  }\href@noop {} {\bibfield  {journal} {\bibinfo  {journal} {Annu. Rev. Fluid
  Mech.}\ }\textbf {\bibinfo {volume} {20}},\ \bibinfo {pages} {111} (\bibinfo
  {year} {1988})}\BibitemShut {NoStop}%
\bibitem [{\citenamefont {Yariv}(2016)}]{Yariv}%
  \BibitemOpen
  \bibfield  {author} {\bibinfo {author} {\bibfnamefont {E.}~\bibnamefont
  {Yariv}},\ }\href@noop {} {\bibfield  {journal} {\bibinfo  {journal} {Phys.
  Rev. Fluids}\ }\textbf {\bibinfo {volume} {1}},\ \bibinfo {pages} {032101(R)}
  (\bibinfo {year} {2016})}\BibitemShut {NoStop}%
\bibitem [{\citenamefont {Varma}\ \emph {et~al.}(2018)\citenamefont {Varma},
  \citenamefont {Montenegro-Johnson},\ and\ \citenamefont {Michelin}}]{Varma}%
  \BibitemOpen
  \bibfield  {author} {\bibinfo {author} {\bibfnamefont {A.}~\bibnamefont
  {Varma}}, \bibinfo {author} {\bibfnamefont {T.}~\bibnamefont
  {Montenegro-Johnson}}, \ and\ \bibinfo {author} {\bibfnamefont
  {S.}~\bibnamefont {Michelin}},\ }\href@noop {} {\bibfield  {journal}
  {\bibinfo  {journal} {Soft Matter}\ }\textbf {\bibinfo {volume} {14}},\
  \bibinfo {pages} {7155} (\bibinfo {year} {2018})}\BibitemShut {NoStop}%
\bibitem [{\citenamefont {Fily}\ and\ \citenamefont {Marchetti}(2012)}]{Fily}%
  \BibitemOpen
  \bibfield  {author} {\bibinfo {author} {\bibfnamefont {Y.}~\bibnamefont
  {Fily}}\ and\ \bibinfo {author} {\bibfnamefont {M.}~\bibnamefont
  {Marchetti}},\ }\href@noop {} {\bibfield  {journal} {\bibinfo  {journal}
  {Phys. Rev. Lett.}\ }\textbf {\bibinfo {volume} {108}},\ \bibinfo {pages}
  {235702} (\bibinfo {year} {2012})}\BibitemShut {NoStop}%
\bibitem [{\citenamefont {Redner}\ \emph {et~al.}(2013)\citenamefont {Redner},
  \citenamefont {Hagan},\ and\ \citenamefont {Baskaran}}]{Redner}%
  \BibitemOpen
  \bibfield  {author} {\bibinfo {author} {\bibfnamefont {G.}~\bibnamefont
  {Redner}}, \bibinfo {author} {\bibfnamefont {M.}~\bibnamefont {Hagan}}, \
  and\ \bibinfo {author} {\bibfnamefont {A.}~\bibnamefont {Baskaran}},\
  }\href@noop {} {\bibfield  {journal} {\bibinfo  {journal} {Phys. Rev. Lett.}\
  }\textbf {\bibinfo {volume} {110}},\ \bibinfo {pages} {055701} (\bibinfo
  {year} {2013})}\BibitemShut {NoStop}%
\bibitem [{\citenamefont {Pohl}\ and\ \citenamefont {Stark}(2014)}]{Pohl1}%
  \BibitemOpen
  \bibfield  {author} {\bibinfo {author} {\bibfnamefont {O.}~\bibnamefont
  {Pohl}}\ and\ \bibinfo {author} {\bibfnamefont {H.}~\bibnamefont {Stark}},\
  }\href@noop {} {\bibfield  {journal} {\bibinfo  {journal} {Phys. Rev. Lett.}\
  }\textbf {\bibinfo {volume} {112}},\ \bibinfo {pages} {238303} (\bibinfo
  {year} {2014})}\BibitemShut {NoStop}%
\bibitem [{\citenamefont {Soto}\ and\ \citenamefont
  {Golestanian}(2014)}]{GolestanianSoto}%
  \BibitemOpen
  \bibfield  {author} {\bibinfo {author} {\bibfnamefont {R.}~\bibnamefont
  {Soto}}\ and\ \bibinfo {author} {\bibfnamefont {R.}~\bibnamefont
  {Golestanian}},\ }\href@noop {} {\bibfield  {journal} {\bibinfo  {journal}
  {Phys. Rev. Lett.}\ }\textbf {\bibinfo {volume} {112}},\ \bibinfo {pages}
  {068301} (\bibinfo {year} {2014})}\BibitemShut {NoStop}%
\bibitem [{\citenamefont {Alarc\'on}\ \emph {et~al.}(2017)\citenamefont
  {Alarc\'on}, \citenamefont {Valeriani},\ and\ \citenamefont
  {Pagonabarraga}}]{Alarcon}%
  \BibitemOpen
  \bibfield  {author} {\bibinfo {author} {\bibfnamefont {F.}~\bibnamefont
  {Alarc\'on}}, \bibinfo {author} {\bibfnamefont {C.}~\bibnamefont
  {Valeriani}}, \ and\ \bibinfo {author} {\bibfnamefont {I.}~\bibnamefont
  {Pagonabarraga}},\ }\href@noop {} {\bibfield  {journal} {\bibinfo  {journal}
  {Soft Matter}\ }\textbf {\bibinfo {volume} {13}},\ \bibinfo {pages} {814}
  (\bibinfo {year} {2017})}\BibitemShut {NoStop}%
\bibitem [{\citenamefont {Voronoj}(1908)}]{Voronoj}%
  \BibitemOpen
  \bibfield  {author} {\bibinfo {author} {\bibfnamefont {G.}~\bibnamefont
  {Voronoj}},\ }\href@noop {} {\bibfield  {journal} {\bibinfo  {journal} {J.
  Reine Angew. Math.}\ }\textbf {\bibinfo {volume} {133}},\ \bibinfo {pages}
  {97} (\bibinfo {year} {1908})}\BibitemShut {NoStop}%
\bibitem [{\citenamefont {Rycroft}\ \emph {et~al.}(2006)\citenamefont
  {Rycroft}, \citenamefont {Grest}, \citenamefont {Landry},\ and\ \citenamefont
  {Bazant}}]{Rycroft}%
  \BibitemOpen
  \bibfield  {author} {\bibinfo {author} {\bibfnamefont {C.}~\bibnamefont
  {Rycroft}}, \bibinfo {author} {\bibfnamefont {G.}~\bibnamefont {Grest}},
  \bibinfo {author} {\bibfnamefont {J.}~\bibnamefont {Landry}}, \ and\ \bibinfo
  {author} {\bibfnamefont {M.}~\bibnamefont {Bazant}},\ }\href@noop {}
  {\bibfield  {journal} {\bibinfo  {journal} {Phys. Rev. E}\ }\textbf {\bibinfo
  {volume} {74}},\ \bibinfo {pages} {021306} (\bibinfo {year}
  {2006})}\BibitemShut {NoStop}%
\bibitem [{Note1()}]{Note1}%
  \BibitemOpen
  \bibinfo {note} {We follow the standard procedure of embedding each particle
  in a $d$-dimensional cell whose $i$-th edge (face) is set to be equally
  distant from the reference particle and its $i$-th nearest
  neighbour.}\BibitemShut {Stop}%
\bibitem [{\citenamefont {Liebchen}\ \emph {et~al.}(2015)\citenamefont
  {Liebchen}, \citenamefont {Marenduzzo}, \citenamefont {Pagonabarraga},\ and\
  \citenamefont {Cates}}]{Benno}%
  \BibitemOpen
  \bibfield  {author} {\bibinfo {author} {\bibfnamefont {B.}~\bibnamefont
  {Liebchen}}, \bibinfo {author} {\bibfnamefont {D.}~\bibnamefont
  {Marenduzzo}}, \bibinfo {author} {\bibfnamefont {I.}~\bibnamefont
  {Pagonabarraga}}, \ and\ \bibinfo {author} {\bibfnamefont {M.}~\bibnamefont
  {Cates}},\ }\href@noop {} {\bibfield  {journal} {\bibinfo  {journal} {Phys.
  Rev. Lett.}\ }\textbf {\bibinfo {volume} {115}},\ \bibinfo {pages} {258301}
  (\bibinfo {year} {2015})}\BibitemShut {NoStop}%
\bibitem [{\citenamefont {Pohl}\ and\ \citenamefont {Stark}(2015)}]{Pohl2}%
  \BibitemOpen
  \bibfield  {author} {\bibinfo {author} {\bibfnamefont {O.}~\bibnamefont
  {Pohl}}\ and\ \bibinfo {author} {\bibfnamefont {H.}~\bibnamefont {Stark}},\
  }\href@noop {} {\bibfield  {journal} {\bibinfo  {journal} {Eur. Phys. J. E}\
  }\textbf {\bibinfo {volume} {38}},\ \bibinfo {pages} {93} (\bibinfo {year}
  {2015})}\BibitemShut {NoStop}%
\bibitem [{Note3()}]{Note3}%
  \BibitemOpen
  \bibinfo {note} {We set such cutoff to the value of $\Lambda = 2R + h$, $h$
  being the hard-core range of interaction.}\BibitemShut {Stop}%
\bibitem [{\citenamefont {Witten}\ and\ \citenamefont {Sander}(1981)}]{Witten}%
  \BibitemOpen
  \bibfield  {author} {\bibinfo {author} {\bibfnamefont {T.}~\bibnamefont
  {Witten}}\ and\ \bibinfo {author} {\bibfnamefont {L.}~\bibnamefont
  {Sander}},\ }\href@noop {} {\bibfield  {journal} {\bibinfo  {journal} {Phys.
  Rev. Lett.}\ }\textbf {\bibinfo {volume} {47}},\ \bibinfo {pages} {140}
  (\bibinfo {year} {1981})}\BibitemShut {NoStop}%
\bibitem [{\citenamefont {Meakin}(1983)}]{MeakinPRL}%
  \BibitemOpen
  \bibfield  {author} {\bibinfo {author} {\bibfnamefont {P.}~\bibnamefont
  {Meakin}},\ }\href@noop {} {\bibfield  {journal} {\bibinfo  {journal} {Phys.
  Rev. Lett.}\ }\textbf {\bibinfo {volume} {51}},\ \bibinfo {pages} {1119}
  (\bibinfo {year} {1983})}\BibitemShut {NoStop}%
\bibitem [{Note4()}]{Note4}%
  \BibitemOpen
  \bibinfo {note} {In principle there can be a dependence on $\mu $ also of the
  fractal dimension $d_f$ ; we assume here, however, that the change in $\mu $
  affects only the characteristic cluster size and not its ``compactness'' (or
  fractality).}\BibitemShut {Stop}%
\bibitem [{\citenamefont {Popescu}\ \emph {et~al.}(2018)\citenamefont
  {Popescu}, \citenamefont {Uspal}, \citenamefont {Eskandari}, \citenamefont
  {Tasinkevych},\ and\ \citenamefont {Dietrich}}]{PopescuNEW}%
  \BibitemOpen
  \bibfield  {author} {\bibinfo {author} {\bibfnamefont {M.}~\bibnamefont
  {Popescu}}, \bibinfo {author} {\bibfnamefont {W.}~\bibnamefont {Uspal}},
  \bibinfo {author} {\bibfnamefont {Z.}~\bibnamefont {Eskandari}}, \bibinfo
  {author} {\bibfnamefont {M.}~\bibnamefont {Tasinkevych}}, \ and\ \bibinfo
  {author} {\bibfnamefont {S.}~\bibnamefont {Dietrich}},\ }\href@noop {}
  {\bibfield  {journal} {\bibinfo  {journal} {Eur. Phys. J. E}\ }\textbf
  {\bibinfo {volume} {41}},\ \bibinfo {pages} {145} (\bibinfo {year}
  {2018})}\BibitemShut {NoStop}%
\bibitem [{\citenamefont {Meakin}(1991)}]{MeakinRevGeo}%
  \BibitemOpen
  \bibfield  {author} {\bibinfo {author} {\bibfnamefont {P.}~\bibnamefont
  {Meakin}},\ }\href@noop {} {\bibfield  {journal} {\bibinfo  {journal} {Rev.
  Geophys.}\ }\textbf {\bibinfo {volume} {29}},\ \bibinfo {pages} {317}
  (\bibinfo {year} {1991})}\BibitemShut {NoStop}%
\bibitem [{\citenamefont {Hansen}\ and\ \citenamefont
  {McDonald}(2006)}]{Hansen}%
  \BibitemOpen
  \bibfield  {author} {\bibinfo {author} {\bibfnamefont {J.-P.}\ \bibnamefont
  {Hansen}}\ and\ \bibinfo {author} {\bibfnamefont {I.}~\bibnamefont
  {McDonald}},\ }\href@noop {} {\emph {\bibinfo {title} {Theory of simple
  liquids}}},\ \bibinfo {edition} {3rd}\ ed.\ (\bibinfo  {publisher}
  {Elsevier},\ \bibinfo {year} {2006})\BibitemShut {NoStop}%
\bibitem [{\citenamefont {Z\"ottl}\ and\ \citenamefont
  {Stark}(2014)}]{ZoettlStark}%
  \BibitemOpen
  \bibfield  {author} {\bibinfo {author} {\bibfnamefont {A.}~\bibnamefont
  {Z\"ottl}}\ and\ \bibinfo {author} {\bibfnamefont {H.}~\bibnamefont
  {Stark}},\ }\href@noop {} {\bibfield  {journal} {\bibinfo  {journal} {Phys.
  Rev. Lett.}\ }\textbf {\bibinfo {volume} {112}},\ \bibinfo {pages} {118101}
  (\bibinfo {year} {2014})}\BibitemShut {NoStop}%
\bibitem [{\citenamefont {Das}\ \emph {et~al.}(2015)\citenamefont {Das},
  \citenamefont {Garg}, \citenamefont {Campbell}, \citenamefont {Howse},
  \citenamefont {Sen}, \citenamefont {Velegol}, \citenamefont {Golestanian},\
  and\ \citenamefont {Ebben}}]{Das}%
  \BibitemOpen
  \bibfield  {author} {\bibinfo {author} {\bibfnamefont {S.}~\bibnamefont
  {Das}}, \bibinfo {author} {\bibfnamefont {A.}~\bibnamefont {Garg}}, \bibinfo
  {author} {\bibfnamefont {A.}~\bibnamefont {Campbell}}, \bibinfo {author}
  {\bibfnamefont {J.}~\bibnamefont {Howse}}, \bibinfo {author} {\bibfnamefont
  {A.}~\bibnamefont {Sen}}, \bibinfo {author} {\bibfnamefont {D.}~\bibnamefont
  {Velegol}}, \bibinfo {author} {\bibfnamefont {R.}~\bibnamefont
  {Golestanian}}, \ and\ \bibinfo {author} {\bibfnamefont {S.}~\bibnamefont
  {Ebben}},\ }\href@noop {} {\bibfield  {journal} {\bibinfo  {journal} {Nat.
  Comm.}\ }\textbf {\bibinfo {volume} {6}},\ \bibinfo {pages} {8999} (\bibinfo
  {year} {2015})}\BibitemShut {NoStop}%
\bibitem [{\citenamefont {Uspal}\ \emph {et~al.}(2015)\citenamefont {Uspal},
  \citenamefont {Popescu}, \citenamefont {Dietrich},\ and\ \citenamefont
  {Tasinkevych}}]{Uspal}%
  \BibitemOpen
  \bibfield  {author} {\bibinfo {author} {\bibfnamefont {W.}~\bibnamefont
  {Uspal}}, \bibinfo {author} {\bibfnamefont {M.}~\bibnamefont {Popescu}},
  \bibinfo {author} {\bibfnamefont {S.}~\bibnamefont {Dietrich}}, \ and\
  \bibinfo {author} {\bibfnamefont {M.}~\bibnamefont {Tasinkevych}},\
  }\href@noop {} {\bibfield  {journal} {\bibinfo  {journal} {Soft Matter}\
  }\textbf {\bibinfo {volume} {11}},\ \bibinfo {pages} {434} (\bibinfo {year}
  {2015})}\BibitemShut {NoStop}%
\bibitem [{\citenamefont {Mozaffari}\ \emph {et~al.}(2016)\citenamefont
  {Mozaffari}, \citenamefont {Sharifi-Mood}, \citenamefont {Koplik},\ and\
  \citenamefont {Maldarelli}}]{Mozaffari}%
  \BibitemOpen
  \bibfield  {author} {\bibinfo {author} {\bibfnamefont {A.}~\bibnamefont
  {Mozaffari}}, \bibinfo {author} {\bibfnamefont {N.}~\bibnamefont
  {Sharifi-Mood}}, \bibinfo {author} {\bibfnamefont {J.}~\bibnamefont
  {Koplik}}, \ and\ \bibinfo {author} {\bibfnamefont {C.}~\bibnamefont
  {Maldarelli}},\ }\href@noop {} {\bibfield  {journal} {\bibinfo  {journal}
  {Phys. Fluids}\ }\textbf {\bibinfo {volume} {28}},\ \bibinfo {pages} {053107}
  (\bibinfo {year} {2016})}\BibitemShut {NoStop}%
\bibitem [{\citenamefont {Saha}\ \emph {et~al.}(2014)\citenamefont {Saha},
  \citenamefont {Golestanian},\ and\ \citenamefont {Ramaswamy}}]{Saha}%
  \BibitemOpen
  \bibfield  {author} {\bibinfo {author} {\bibfnamefont {S.}~\bibnamefont
  {Saha}}, \bibinfo {author} {\bibfnamefont {R.}~\bibnamefont {Golestanian}}, \
  and\ \bibinfo {author} {\bibfnamefont {S.}~\bibnamefont {Ramaswamy}},\
  }\href@noop {} {\bibfield  {journal} {\bibinfo  {journal} {Phys. Rev. E}\
  }\textbf {\bibinfo {volume} {89}},\ \bibinfo {pages} {062316} (\bibinfo
  {year} {2014})}\BibitemShut {NoStop}%
\bibitem [{\citenamefont {Bickel}\ \emph {et~al.}(2014)\citenamefont {Bickel},
  \citenamefont {Zecua},\ and\ \citenamefont {Wurger}}]{Bickel}%
  \BibitemOpen
  \bibfield  {author} {\bibinfo {author} {\bibfnamefont {T.}~\bibnamefont
  {Bickel}}, \bibinfo {author} {\bibfnamefont {G.}~\bibnamefont {Zecua}}, \
  and\ \bibinfo {author} {\bibfnamefont {A.}~\bibnamefont {Wurger}},\
  }\href@noop {} {\bibfield  {journal} {\bibinfo  {journal} {Phys. Rev. E}\
  }\textbf {\bibinfo {volume} {89}},\ \bibinfo {pages} {050303(R)} (\bibinfo
  {year} {2014})}\BibitemShut {NoStop}%
\bibitem [{\citenamefont {Leibchen}\ \emph {et~al.}(2017)\citenamefont
  {Leibchen}, \citenamefont {Marenduzzo},\ and\ \citenamefont
  {Cates}}]{Liebchen2017}%
  \BibitemOpen
  \bibfield  {author} {\bibinfo {author} {\bibfnamefont {B.}~\bibnamefont
  {Leibchen}}, \bibinfo {author} {\bibfnamefont {D.}~\bibnamefont
  {Marenduzzo}}, \ and\ \bibinfo {author} {\bibfnamefont {M.}~\bibnamefont
  {Cates}},\ }\href@noop {} {\bibfield  {journal} {\bibinfo  {journal} {Phys.
  Rev. Lett.}\ }\textbf {\bibinfo {volume} {118}},\ \bibinfo {pages} {268001}
  (\bibinfo {year} {2017})}\BibitemShut {NoStop}%
\end{thebibliography}%

\end{document}